\documentclass[reqno]{amsart}

\usepackage{amsmath,amssymb,amsthm,amsfonts}
\usepackage{color}
\usepackage{ulem}
\usepackage{hyperref}
\usepackage{a4wide}
\usepackage{mathrsfs} 

\numberwithin{equation}{section}

\newtheorem{theorem}{Theorem}[section]
\newtheorem{lemma}[theorem]{Lemma}
\newtheorem{prop}[theorem] {Proposition}

\newtheorem{definition}[theorem] {Definition}

\theoremstyle{definition}

\theoremstyle{remark}
\newtheorem{remark}[theorem]{Remark}
\newtheorem{example}[theorem]{Example}

\newcommand{\e}{\mathrm{e}} 
\newcommand{\N}{\mathbb{N}}
\newcommand{\R}{\mathbb{R}}

\newcommand{\C}{\mathbb{C}}

\newcommand{\dd}{\mathrm{d}} 

\newcommand{\vect}[1]{\boldsymbol{#1}}

\newcommand{\be}{\begin{equation}}
\newcommand{\ee}{\end{equation}}
\newcommand{\ba}{\begin{equation} \begin{aligned}}
\newcommand{\ea}{\end{aligned}\end{equation}}
\newcommand{\bes}{\begin{equation*}}
\newcommand{\ees}{\end{equation*}}

\def\1{{\mathchoice {1\mskip-4mu\mathrm l}      
{1\mskip-4mu\mathrm l}
{1\mskip-4.5mu\mathrm l} {1\mskip-5mu\mathrm l}}}

\begin{document}

\title{Cluster expansions, trees, inversions and correlations}
\author{ Dimitrios Tsagkarogiannis}
\address{Dipartimento di Ingegneria e Scienze dell'Informazione e Matematica, Universit\`a degli Studi dell'Aquila, 67100 L'Aquila, Italy}
\email{dimitrios.tsagkarogiannis@univaq.it}

\date{}
\begin{abstract} 
	We review some recent progress on applications of Cluster Expansions. We focus on a system of classical particles living in a continuous medium and interacting via a stable and tempered pair potential. We review the cluster expansion in both the canonical and the grand canonical ensemble and compute thermodynamic quantities such as the pressure, the free energy as well as various correlation functions. We derive the equation of state either by performing inversion of the density-activity series or directly in the canonical ensemble. Further applications to the liquid state expansions and the relevant closures are discussed, in particular their convergence in the gas regime. 
\\
	
	{\it Dedicated to Errico Presutti for his 80th birthday, with deep gratitude.}
\\


\end{abstract}

\maketitle

\tableofcontents

\section{Introduction}

Deriving thermodynamic quantities from microscopic models based on physical principles is one of the main challenges of both theoretical and computational methods in statistical mechanics. The inverse question is equally intriguing: based on experimental
or computational data is it possible to design the atomistic parameters of the system such as the interaction potential?

In the late 30's J.~E. Mayer and collaborators \cite{M37, MA, MH, HM, M39, mayerbook} suggested a theory for expressing thermodynamic quantities in terms of power series of their parameters
(activity or density) for non-ideal gases.
More results on correlation functions appeared immediately after, \cite{K39, MMo, BG, KS}.
Their range of validity was initially debated \cite{BF38, UK, Tem54}, until the first rigorous results appeared in the high-temperature and low-density
regime: first in \cite{BH} based on a fixed-point argument for the Kirkwood-Salsburg equations and later
with the proof of the convergence of the virial expansion by Groeneveld in 1963 \cite{g67} and Lebowitz and Penrose in 1964~\cite{lebowitz-penrose1964virial}, building on the previously established convergence of the activity expansion of the pressure by using again the Kirkwood-Salsburg integral equation \cite{penrose1963, Rcor} or by establishing the tree-graph inequality \cite{P}. 
Since then, many power series expansions between different thermodynamic quantities have been established \cite{morita-hiroike1, morita-hiroike3, DD, stell1964}, usually without checking their radius of convergence.
Since their complexity increases dramatically to higher orders, several empirical closures have been suggested and tested with experimental and computational data, even in
denser regimes such as the liquid state. As a result, by today this research has been 
registered as a theory for the liquid state \cite{HMD}, \cite{santos} which works well in some cases but with still
much to discover. Furthermore, 
new techniques have been developed mainly exploiting the dramatic increase 
of computational power.
However, for more complex systems these computational methods still require very 
long running times, so maybe it would be fruitful to re-investigate the analytic methods 
and based on new intuition from analysis, probability and combinatorics
eventually suggest new computational methods.

Back to the rigorous results, the main steps of these expansions are the following:
we first obtain an expansion of the pressure with respect to the activity (``Mayer's first theorem", \cite{UF}) exploiting the representation of the 
grand canonical ensemble as a generating
function of some combinatorial species (simple graphs), see 
Section~\ref{sec:species}. A similar expansion holds for the density as well.
The convergence of these power series is possible thanks to the tree-graph inequality given in Section~\ref{sec:conv} where we also comment on some recent progress.
The next step is to invert the density-activity expansion by using tree structures 
and this is presented in
Section~\ref{sec:inv} where a more general framework to study inhomogeneous systems as well as correlation functions is given.
In order to obtain the equation of state one needs to plug the resulting expansion of the activity as a function of the density into the pressure-activity expansion and re-sum obtaining
also a bound for the radius of convergence of the composed power series combining the convergence results for the inversion and for the activity expansion. This is the context of
``Mayer's second theorem" which we present in Section~\ref{sec:M2}.
Alternatively, in order to derive the equation of state, being a density expansion, one could work directly in the canonical ensemble. This is presented in Section~\ref{sec:canonical} where we also comment on the comparison with the inversion.
Last but not least, in Section~\ref{sec:correlations} we repeat the above procedure for various
correlation functions in both the canonical and the grand-canonical ensemble. We conclude
with a brief comment on more involved inversions of the second correlation function with respect to the pair inter-particle potential as well as with a short discussion about closures.

This is a review article and we skip the detailed proofs. Whenever relevant we hint on the main ideas and guide the reader to the corresponding literature for the full proofs.
Summarizing, the main points of this contribution are the following:
\begin{enumerate}
\item Present various expansions of thermodynamic quantities and connect them to the underlying combinatorial structures.
\item Provide inversion theorems for the rigorous justification of the density expansions in the gas regime.
\item Show that the various density expansions can be established not only in the grand-canonical ensemble using inversions, but also in the canonical ensemble in a direct way. 
\item Investigate how one could turn a graphical demonstration of an expansion or of an inversion into a rigorous proof about its radius of convergence.
\end{enumerate}


\section{The model}

Let $\mathbb X$ be a measurable space and $V:\mathbb X\times \mathbb X\to \mathbb R\cup \{\infty\}$ a measurable pair potential which we choose to be symmetric: $V(x,y) = V(y,x)$. We assume that for some measurable function $B:\mathbb X\to [0,\infty)$, we have the stability condition 
\begin{equation}\label{stability}
	\sum_{1\leq i<j\leq n} V(x_i, x_j)\geq -\sum_{i=1}^n B(x_i),
\end{equation}
for all $n\geq 2$ and $x_1,\ldots, x_n\in \mathbb X$. 
Define the Hamiltonian:
$$
	H_n(x_1,\ldots, x_n):= \sum_{1\leq i<j\leq n} V(x_i, x_j),
$$
for $n\geq 2$ and $H_0=0$, $H_1=0$. 
For simplicity of the presentation we consider the case $\mathbb X=\Lambda\subset\mathbb R^d$ and later (after equation \eqref{activity}) we will return
to the more general set-up.
The canonical ensemble consists of a probability measure on the configuration space $\Lambda^N$ of $N$ indistinguishable particles.
It is given by the (canonical) Gibbs measure 
\begin{equation}\label{cGm}
\mu_{\Lambda, N, \beta}(\dd\mathbf x):=\frac{1}{Z_{\Lambda,N,\beta}}\e^{-\beta H_N(\mathbf x)}\frac{\dd\mathbf x}{N!}.
\end{equation}
The normalization $Z_{\Lambda,N,\beta}$ is called canonical partition function and it is given by
\begin{equation}\label{can_part_fcn}
Z_{\Lambda,N,\beta}:=\frac{1}{N!}\int_{\Lambda^N}\e^{-\beta  H_N(\vect x)}\dd \vect x.
\end{equation}
Alternatively, one can also consider the grand canonical ensemble on the space 
$\oplus_{N=0}^{\infty}\Lambda^N$ with the grand canonical measure $\mu^{\text{g.c.}}_{\Lambda,\beta}(z)$ being described by its marginals when $N$ particles are fixed, which are given by
$\frac{z^N}{\Xi_{\Lambda,\beta}(z)}\e^{-\beta H_N(\mathbf x)}\frac{\dd\mathbf x}{N!}
$, for some control parameter $z\in\mathbb R$ called {\it activity}.
The corresponding normalization, the grand-canonical partition function, is given by:
\begin{equation}\label{gcan_part_fcn}
\Xi_{\Lambda,\beta}(z):=\sum_{N=0}^{\infty}z^N Z_{\Lambda,N,\beta}.
\end{equation}
Note that all relevant 
properties of the corresponding macroscopic system are registered in the partition function.
For example,
the {\it Helmholtz free energy at finite volume} is given by
\begin{equation}\label{fe}
F_{\beta,\Lambda}(N):=-\frac{1}{\beta|\Lambda|}\ln Z_{\Lambda,N,\beta}
\end{equation}
and the {\it pressure at finite volume}, by:
\begin{equation}\label{p}
P_{\beta,\Lambda}(z):=\frac{1}{\beta|\Lambda|}\ln \Xi_{\Lambda,\beta}(z).
\end{equation}
The goal of statistical mechanics is to compute the above quantities (among others) and study their dependence as we vary the parameters. 
Of particular interest is to understand if they exhibit singularities
as we pass to the thermodynamic limit which consists of sending $\Lambda\uparrow \mathbb R^d$ (for the grand canonical ensemble) and also $N\to\infty$ such that $\frac{N}{|\Lambda|}\to\rho$, for some $\rho>0$ (for the canonical). This is a hard question and in this note
we will content ourselves by understanding when this does not happen, i.e., when
it can be proved that they are analytic functions with respect to the parameters.

\section{Combinatorial species and graph generating functions}\label{sec:species}

The goal is to write \eqref{fe} and \eqref{p} as absolutely convergent power series
with respect to the relevant parameters.
As mentioned in the introduction, this idea originates in the works of J. E. Mayer and collaborators (also inspired by some previous works such as \cite{U, Y})
and it is easier to demonstrate it working with the
grand-canonical partition function as it is explained below.
In fact, letting $\mathcal G$ be the combinatorial species of labelled {\it simple graphs}, we
write $\mathcal G(z)$ for the corresponding generating function of labelled structures and $\mathcal G_V$ for the collection of all {\it simple graphs} on the set $V\subset [N]:=\{1,\ldots,N\}$. With a slight abuse of notation we will also denote by $\mathcal G_N$ 
the set of $\mathcal G$ structures on $N$ vertices, whenever there is no risk of confusion.
We can also consider that each graph comes with a {\it weight} $w:\mathcal G\to\mathbb R$.
Then \eqref{gcan_part_fcn}
can be written as
\begin{equation}\label{xig}
\Xi_{\Lambda}(z)=\mathcal G(z):=\sum_{N=0}^{\infty}\frac{z^N}{N!}G_N,\qquad G_N:=\sum_{g\in\mathcal G_{[N]}}w_{\Lambda}(g),
\end{equation}
where we have removed from the notation the dependence on the inverse temperature $\beta$ as it will not be studied here and
where $G_N$ is the (weighted) number of graphs of cardinality $N$ with weight:
\begin{equation}\label{weight}
w_{\Lambda}(g):=\int_{\Lambda^{|g|}} \prod_{\{i,j\}\in E(g)}f(x_i,x_j)\,\dd \mathbf x, \quad \mathrm{where}\quad f(x_i,x_j):=\e^{-\beta V(x_i,x_j)}-1.
\end{equation}
Note also that we neglected from the notation the dependence of $\mathcal G$ and $G_N$ on the volume $\Lambda$ which is now ``hidden" in the weight function $w_{\Lambda}$.
We denote by $|g|$ the cardinality of the vertices of the graph $g$ and by $E(g)$ 
the set of edges of the graph $g$.
Similarly, we will later denote by $\mathcal C_n$ the set of all {\it connected graphs} on $n$ vertices and by $\mathcal C(z)$ the weighted generating function of the combinatorial species of connected graphs, i.e., $\mathcal C(z):=\sum_{n\geq 1}\frac{z^n}{n!}C_n$
with $C_n:=\sum_{g\in\mathcal C_N}w_{\Lambda}(g)$.
At this point, we give a couple more definitions: 
\begin{itemize}
\item a {\it cutpoint} (or articulation point) of a connected graph $g$ is a vertex of $g$ 
whose removal yields a disconnected graph. 
\item A connected graph is called {\it 2-connected} 
if it has no cutpoint. 
\item A {\it block} in a simple graph is a maximal 2-connected subgraph.
\end{itemize}
Hence, a connected graph can be viewed as a graph whose blocks are 2-connected.
We denote by $\mathcal B_n$ the set of {\it 2-connected graphs} on $n$ vertices.
For a more systematic presentation of the combinatorial structures we refer to 
\cite{bergeron-labelle-leroux1998book, leroux2004} (see also \cite{tate}).
In particular, it is easy to see that a simple graph is the disjoint union of connected graphs
which in terms of generating functions it gives:
\begin{equation}\label{csexp}
P_{\Lambda}(z)=\frac{1}{|\Lambda|}\ln\mathcal G(z)=\frac{1}{|\Lambda|}\mathcal C(z).
\end{equation}
Note that \eqref{csexp} can be re-written in the more familiar form as in \cite{poghosyan-ueltschi2009}:
\begin{eqnarray}\label{cexp2}
P_{\Lambda}(z) & = &
\frac{1}{|\Lambda|}\ln\left(
1+\sum_{N=1}^\infty \frac{z^N}{N!}\int_{\Lambda^N}\phi(x_1,\ldots,x_N)\dd \vect x\right)\nonumber\\
	& = & \frac{1}{|\Lambda|}\sum_{n=1}^\infty \frac{z^n}{n!}\int_{\Lambda^n}\phi^\mathsf T(x_1,\ldots,x_n)\dd \vect x\nonumber\\
& = & \sum_{n=1}^{\infty}\frac{z^n}{n!}\frac{1}{|\Lambda|}\sum_{g\in\mathcal C_n}w_{\Lambda}(g)=:\sum_{n=1}^{\infty}b_n z^n,
\end{eqnarray}
where for a generic binary map $h:\mathbb X\times\mathbb X\to\mathbb R$ we have introduced the coefficients:
\begin{equation}\label{phi}
\phi(x_1,\ldots,x_n):=\prod_{1\leq i < j\leq n}(1+h(x_i,x_j))=\sum_{g\in\mathcal G_n}
\prod_{\{i,j\}\in E(g)} h(x_i,x_j)
\end{equation}
and 
\begin{equation}\label{phit}
\phi^\mathsf T(x_1,\ldots,x_n):=\sum_{g\in\mathcal C_n}\prod_{\{i,j\}\in E(g)}h(x_i,x_j).
\end{equation}
Here we have chosen $h=f$ as in \eqref{weight}, but we gave the definitions of $\phi$
and $\phi^\mathsf T$ for a generic $h$ as other choices will be relevant later (see Section~\ref{sec:canonical}).
Furthermore, as mentioned before \eqref{csexp}, we note that
\begin{equation}\label{part_phi}
\phi(x_1,\ldots,x_n)=
\sum_{k=1}^n
\sum_{\substack{\{P_1,\ldots,P_k\}\\ \in\Pi_k(1,\ldots,n)}}\prod_{i=1}^k \phi^\mathsf T(\mathbf x_{P_i}),
\end{equation}
where $\Pi_k(1,\ldots,n)$ is the set of partitions of $\{1,\ldots,n\}$ into $k$ blocks. 
Then for a given partition $\{P_1,\ldots, P_k\}\in\Pi_k(1,\ldots,n)$ and a given block
$P_i=\{j_1,\ldots, j_{|P_i|}\}\subset\{1,\ldots,n\}$ we set $\mathbf x_{P_i}:=(x_{j_1},\ldots,x_{j_{|P_i|}})$

The next important issue is to investigate for which values of $z$ the above series is absolutely convergent.

\section{Convergence}\label{sec:conv}

The main idea is to estimate the sum over connected graphs in \eqref{cexp2} by a sum over trees. This is the context of the following tree-graph inequality:

\begin{lemma}\label{lemTGI}
If \eqref{stability} holds, then
\begin{equation}\label{tgi}
\left|\sum_{g\in\mathcal C_n}\prod_{\{i,j\}\in E(g)} f(x_i,x_j)\right|
\leq
\e^{\sum_{i=1}^n B(x_i)}\sum_{\tau\in\mathcal T_n}\prod_{\{i,j\}\in E(g)} \bar f(x_i,x_j),
\end{equation}
where
\begin{equation}\label{barf}
\bar f(x_i,x_j):=1 - \e^{-\beta|V(x_i,x_j)|}
\end{equation}
and $\mathcal T_n$ is the set of trees on $n$ vertices.
\end{lemma}

For the proof of the recent version with $\bar f$ as in \eqref{barf} we refer to Procacci and Yuhjtman~\cite{procacci-yuhjtman2017}, see also~\cite{ueltschi2017, procacci2017correction}. For earlier versions we refer to \cite{P, basuev, Br, MalyshevMinlos1991}. In fact, there is an interesting connection between Banach fixed point argument for correlation functions and tree identities \cite{BH, P, ruelle1969book, BF, MalyshevMinlos1991, MP, kuna, kuna2001, poghosyan-ueltschi2009}.
With this lemma we have the following consequence for \eqref{cexp2}:
\begin{eqnarray}\label{cons}
\sum_{n\geq 1}z^n
\frac{1}{n!}\frac{1}{|\Lambda|}\left|\sum_{g\in\mathcal C_n}w_{\Lambda}(g)\right| & \leq &
\frac{1}{|\Lambda|}\sum_{n\geq 1}\frac{z^n}{n!}\int_{\Lambda^n}
\e^{\sum_{i=1}^n B(x_i)}
\sum_{\tau\in\mathcal T_n}\prod_{\{i,j\}\in E(g)} \bar f(x_i,x_j)\,\dd\mathbf x\nonumber\\
& =:  &
\frac{1}{|\Lambda|}\int_{\Lambda}T^{\circ}_{x_0}(z) z\e^{B(x_0)}\,\dd x_0.
\end{eqnarray}
Notice that the quantity $T^{\circ}_{x_0}(z)$ - the weighted
generating function of the combinatorial species of simple trees rooted at $x_0$ -
satisfies the equation:
\begin{equation}\label{tree_eq}
T^{\circ}_{x_0}(z)=\exp\left\{
\int_{\Lambda}\bar f(x_0,x) \, T^{\circ}_{x}(z) \, z\e^{B(x)}\dd x
\right\},
\end{equation}
see Faris \cite{faris}, Section 3.1.

With the above, ``Mayer's first theorem" of absolute convergence of the power
series expression of the pressure with respect to the activity follows from the following proposition:

\begin{prop}\label{propM1}
For every $x_0\in\Lambda$, $T^{\circ}_{x_0}(z)<\infty$ if and only if for some positive function
$a:\Lambda\subset\mathbb R^d\to\mathbb R_+$ and all $x_0$:
\begin{equation}\label{cond}
\int_{\Lambda}\bar f(x_0,x)\e^{a(x)}z\e^{B(x)}\, \dd x \leq a(x_0).
\end{equation}
\end{prop}
The proof is given by induction and for the details we refer to \cite{poghosyan-ueltschi2009}, Theorem 2.1 and \cite{jansen2019pointproc}, Proposition 2.1.

\section{Inversion}\label{sec:inv}

Another thermodynamic quantity of interest is the density which can be defined
as the expected value of observing a given fraction of particles within a box 
$\Lambda\subset\mathbb R^d$ at a given activity $z$.
Let us consider the sample space $\Omega_{\Lambda}:=\oplus_{N=0}^{\infty}\Lambda^N$
(with the proper $\sigma$-algebra)
and the random variable $N_{\Lambda}:=|\mathbf x\cap \Lambda|$.
By computing the first moment with respect to the grand-canonical measure $\mu^{\text{g.c.}}_{\Lambda}(z)$ we define:
\begin{equation}\label{density}
\rho_{\Lambda}(z) := \mathbb E_{\mu^{\text{g.c.}}_{\Lambda}(z)}\left[\frac{N_{\Lambda}}{|\Lambda|}\right]
=z\frac{d}{dz}P_{\Lambda}(z).
\end{equation}
Referring again to \cite{leroux2004} this can be viewed as the generating function of ``rooted" connected graphs, denoted by $\rho_{\Lambda}(z)=\mathcal C^{\bullet}(z)$.
Given a species, e.g. the connected graphs $\mathcal C$ (similarly for the other species $\mathcal B$, $\mathcal T$), the operation of rooting (or pointing) $\mathcal C\mapsto\mathcal C^{\bullet}$
at an element of the underlying set can be defined by
\begin{equation}\label{rooting}
\mathcal C^{\bullet}(z):=z\mathcal C'(z),
\end{equation}
where the derivative $\mathcal C'(z)$ is defined as adding to the structure an external (unlabelled) element $*$; hence, if we choose the underlying set to be $[n]$, we have: 
$\mathcal C'_{[n]}=\mathcal C_{[n]\cup\{*\}}$.
Comparing \eqref{rooting} to the definition \eqref{cons}
we note that we can define various types of generating functions of rooted species by specifying or not the label of the root and by multiplying or not with $z$ ($\mathcal C^{\bullet}$ vs $\mathcal C^{\circ}$).
The goal of this section is to invert the above formula and plug it into the pressure in order
to obtain an equation between the pressure and the density, known as equation of state.
Apart of its relevance in applications there is an interesting mathematical question about
the convergence of these series which
we describe with an example:

\subsection{A simple example}

\begin{example}\label{ex1}
Let $P(z)$ be the generating function of ``cyclic permutations", i.e.,
\begin{equation}\label{perm}
P(z)=\sum_{n\geq 0}\frac{z^n}{n!}n!=\frac{1}{1-z},\qquad |z|<1.
\end{equation}
It is easy to see that it is the exponential (generating function of the combinatorial species ``sets") of the generating function of ``cycles", given below:
\begin{equation}\label{cycles}
C(z)=\sum_{n\geq 1}\frac{z^n}{n!}(n-1)!=-\ln(1-z), \qquad 0<z<1.
\end{equation}
Suppose that we consider a particular gas whose pressure is given by $C(z)$ as in \eqref{cycles}, valid for $0<z<1$, i.e., there is a singularity at $z=1$. 
The corresponding density from \eqref{density} is given by:
\begin{equation}\label{sd}
\rho(z)=z C'(z)=\frac{z}{1-z}\Leftrightarrow z=\frac{\rho}{1+\rho}.
\end{equation}
Note that due to the explicit formula the inversion is easy.
Then, substituting back to \eqref{cycles} we obtain the following equation of state:
\begin{equation}\label{seos}
\bar C(\rho):=C(z(\rho))=-\ln(1-\frac{\rho}{1+\rho})=\ln(1+\rho).
\end{equation}
Observe that the latter is valid for all values of $\rho>0$, i.e., the singularity at $z=1$ for
the pressure is not present anymore when the pressure is expressed in terms of the density. 
This might indicate that probably it is more useful to view the ``pressure" $C$ as a function of $\rho$ rather than a function of $z$.
Is this true for the general model? That is, does a similar ``direct" expansion with respect to the density for the general model enjoy of a similar property? 
\end{example}

\subsection{Inversion with trees}\label{sec: invtrees}

Before investigating the convergence of the inverse series, we give a strategy on how to perform the inversion.
From \eqref{density} we have:
\begin{equation}\label{inv}
\rho_{\Lambda}(z)  =  zP_{\Lambda}'(z) =
z\left(1+ \sum_{n=1}^\infty \frac{z^n}{n!}\int_{\Lambda^n}\phi^\mathsf T(0, x_1,\ldots,x_n)\dd \vect x \right)
= z \e^{-A(0;z)},
\end{equation}
where $A(0;z)$ is given by the following formula by considering $q\equiv 0\in\Lambda$:
\begin{equation}\label{A}
A(q;z):=\sum_{n\geq 1}\frac{z^n}{n!}\int_{\Lambda^n}A_n(q;x_1\ldots x_n)\,\dd x_1\ldots \dd x_n
\end{equation}
and (in our case)
\begin{equation}\label{A1}
	A_n(q; x_1,\ldots,x_n):=-\left[\prod_{j=1}^n (1+f(q,x_j))-1\right]
	\sum_{g\in\mathcal C_n}\prod_{\{i,j\}\in E(g)}f(x_i, x_j).
\end{equation}
Note that similarly to \eqref{csexp}, in \eqref{inv} we viewed $\mathcal C^{\bullet}$ 
as the following operations: first take the combinatorial species of {\it sets} over all vertices except the root $0$ (giving the exponential), then sum over all connected graphs within each member of the sets (the second factor of $A_n$ in \eqref{A1}) together with all possible links to the root $0$ (first factor of $A_n$).
The goal is to find a solution $z(\rho_\Lambda)=\rho_\Lambda \bar T_0^\circ(\rho_\Lambda)$ of \eqref{inv} expressed in terms of a power series $\bar T_0^\circ$ to be found. Hence, we rewrite \eqref{inv} as the following {\it fixed point} problem:
\begin{equation}\label{FP}
\bar T_0^\circ(\rho_\Lambda)=\e^{A(0;\rho_\Lambda \bar T_0^\circ(\rho_\Lambda))},
\end{equation}
or, more explicitly, as:
\be \label{tree-eq}
	\bar T_0^\circ(\rho) = \exp\Biggl( \sum_{n=1}^\infty \frac{1}{n!} \int_{\mathbb X^n} A_n(0;x_1,\ldots, x_n) \bar T_{x_1}^\circ(\rho)\cdots \bar T_{x_n}^\circ(\rho) \rho^n \dd x_1\cdots \dd x_n\Biggr). 
\ee 
Note that equation~\eqref{tree-eq} is understood in the sense of formal power series. 
Once one can prove absolute convergence in the sense of Proposition~\ref{propM1},
then equation~\eqref{tree-eq} turns into a relation of analytic functions.
Here we give the main ideas of the inversion and we will present the rigorous statements in Section~\ref{sec:one_point}.

Comparing \eqref{tree-eq} with \eqref{tree_eq} we observe a similar iterative structure, hence such a solution might be expressed as a
power series of a special class of trees to be determined.
This procedure of inverting using trees is by now a well-understood strategy which dates back to \cite{gallavotti}
and it appears in many contexts. See \cite{JKT2} for a more detailed discussion about the combinatorics and other possible applications.
Indeed, in \cite{JKT_JFA}, Proposition 2.6, it is proved that the unique solution of \eqref{inv} is given by
$z=\rho_\Lambda \bar T_0^\circ(\rho_\Lambda)$ where $\bar T_0^\circ$ (the unique solution of \eqref{tree-eq}) is a generating function of some special type of trees given below.
Consider a genealogical tree that keeps track not only of mother-child relations, but also of groups of  siblings born at the same time. This results in a tree for which children of a vertex are partitioned into cliques (singletons, twins, triplets, etc.). Accordingly for $n\in \N$ we define $\mathcal{TP}_n^\circ$ as the set of pairs $(T, (P_i)_{0\leq i \leq n})$ consisting of: 
\begin{itemize}
	\item A tree $T$ with vertex set $[n]:=\{0,1,\ldots,n\}$. The tree is considered rooted in $0$ (the ancestor).
	\item For each vertex $i\in \{0,1,\ldots,n\}$, a set  partition $ P_i$ of the set of children\footnote{The members of the partition are assumed to be non-empty, except we consider the partition of the empty set.} of $i$. If $i$ is a leaf (has no children), then we set $P_i = \varnothing$. 
\end{itemize} 
For $x_0,\ldots, x_n\in \Lambda$, we define the weight of an enriched tree $(T, (P_i)_{0\leq i \leq n})\in \mathcal{TP}_n^\circ$ as 
\be \label{eq:treeweight}
	w\bigl( T,(P_i)_{0\leq i \leq n}; x_0,x_1,\ldots, x_n\bigr)  := \prod_{i=0}^n \prod_{J\in P_i} A_{\#J +1}\bigl(x_i;(x_j)_{j\in J}\bigr),
\ee
with $\prod_{J\in \varnothing} =1$. So the weight of an enriched tree is a product over all cliques of twins, triplets, etc., contributing each a weight that depends on the variables $x_j$ of the clique members and the variable $x_i$ of the parent. 
With the above we have that the family of power series $(\bar T_q^\circ)_{q\in\mathbb X}$ which satisfies \eqref{tree-eq} is given by 
	\be\label{specialtrees}
		\bar T_0^\circ(\rho) = 1+ \sum_{n=1}^\infty \frac{1}{n!} \int_{\mathbb X^n} \sum_{(T,(P_i)_{i=0,\ldots,n})\in \mathcal{TP}_n^\circ} 
		w\bigl( T,(P_i)_{i=0,\ldots,n}; 0,x_1,\ldots, x_n\bigr) \rho^n \dd \vect x. 
	\ee

\begin{remark}\label{rem:conv}
Comparing
to Example~\ref{ex1} we have that $C'(z)=e^{C(z)}$ or $A(0;z)=C(z)$, that is
\begin{equation}\label{extrees}
T(\rho)=\frac{\rho}{1+\rho}=\rho\sum_{n\geq 0}(-\rho)^{n},
\end{equation}
for which we need $0<\rho<1$. Observe that both \eqref{extrees} 
and \eqref{seos} have a pole at $-1$, but no restriction for positive values, which one would like to exploit while establishing the region of analyticity.
Furthermore, when \eqref{extrees} is viewed as a power series
expansion (originating from the tree expansion solution of the inversion formula) its absolute convergence is valid within the radius of convergence $\rho<1$, while the direct calculation and consequently the final formula \eqref{seos} does not have that constraint.
\end{remark}

\begin{remark}
The second observation is that by composing $A$ with $\bar T^{\circ}_0$ in \eqref{FP}
one obtains some
simplifications.
From \cite{leroux2004}, Figure 4, \cite{stell1964}, Section 5, relation (5-6)  and \cite{morita-hiroike3}, formula (4.4) we have the following
diagrammatic construction:
a connected graph with root $0$ 
can be viewed 
as a partition (the exponential) of the following structure: in each element of the partition,
we start with the root $0$ 
which is part of a block (a rooted 2-connected component). Then, from each vertex of the 2-connected component it can start a new connected graph having that vertex as a root.
In formulas:
\begin{equation}\label{invfor_comb}
\mathcal C^{\bullet}(z)=z\cdot\exp\{\mathcal B'(\mathcal C^{\bullet}(z))\},
\end{equation}
or equivalently,
\begin{equation}\label{invfor}
\rho_{\Lambda}(z)
=z \e^{\mathcal B'(\rho_{\Lambda}(z))},
\end{equation}
where $\mathcal B$ is the generating function of the 2-connected graphs:
\begin{equation}\label{twoconn}
\mathcal B(\rho)=\sum_{n=2}^{\infty}B_n\frac{\rho^{n}}{n!} , \qquad B_n:=\sum_{g\in\mathcal B_n}w_{\Lambda}(g),
\end{equation}
(where again we do not explicit the dependence on $\Lambda$ in the notation $B_n$ and $\mathcal B$).
Note that in the literature it is usually called the ``irreducible" coefficient 
\begin{equation}\label{betan}
\beta_n:=\frac{1}{n!}\frac{1}{|\Lambda|}B_{n+1}.
\end{equation}
Thus, in view of \eqref{invfor_comb} the term $A(0;z)$ in \eqref{inv}
can be written as
\begin{equation}\label{stellandco}
- A(0;z) = \sum_{n=1}^\infty \frac{1}{n!}\int_{\mathbb X^n} D_{n+1}(0,x_1,\ldots,x_n)\prod_{i=1}^n \e^{- A(x_i;z)} z^n(\dd \vect x),
\end{equation} 
where 
\begin{equation}\label{twoconn_coeff}
D_n(x_1,\ldots,x_n):=\sum_{g\in\mathcal B_n}\prod_{\{i,j\}\in E(g)}f(x_i,x_j)
\end{equation}
and $\mathcal B_n$ is the set of 2-connected graphs on $n$ vertices.
By plugging in $z=\rho_{\Lambda}\bar T^{\circ}_0(\rho_{\Lambda})$ we obtain:
\begin{equation}\label{comp}
A(0;\rho_{\Lambda}\bar T^{\circ}_0(\rho_{\Lambda}))=
-\sum_{n=1}^{\infty}\frac{1}{n!}\int_{\Lambda^n} D_{n+1}(0,x_1,\ldots,x_n)\rho_{\Lambda}^n\dd x_1\ldots \dd x_n.
\end{equation}
Thus, \eqref{comp} can be viewed as a ``constructive" way of obtaining \eqref{invfor} and at the same time proving that the resulting series over 2-connected graphs is absolutely convergent:
one first solves the fixed point problem \eqref{FP} obtaining $\bar T^{\circ}_0$ and then composes it with $A(0;\cdot)$. 
The convergence is established as the composition of two absolutely convergent series. For the detailed proof see Theorem 3.5 in \cite{JKT_JFA} and
we will come back to this observation in Section~\ref{sec:one_point}.
See also \cite{J} for an alternative way of proving the convergence of \eqref{comp}
by revisiting Groeneveld's proof \cite{g67} based on recurrence relations for graph weights related to the Kirkwood-Salsburg integral equation for correlation functions.
\end{remark}

\section{Mayer's second theorem}\label{sec:M2}

The Helmholtz free energy \eqref{fe} and the pressure \eqref{p} are related by a Legendre transform. In fact, at finite volume $\Lambda$ considering that the number of particles can not
exceed $\rho_{\max}|\Lambda|$ we have:
\begin{equation}\label{LD}
P_{\Lambda}(z)=\frac{1}{\beta|\Lambda|}\ln
\sum_{N=0}^{\rho_{\max}|\Lambda|}z^N\e^{-\beta|\Lambda| F_{\Lambda}(N)}
\leq \sup_{N\leq\rho_{\max}|\Lambda|}\{\frac{N}{\beta|\Lambda|}\ln z-F_{\Lambda}(N)\}
+o_{|\Lambda|}(1),
\end{equation}
for large $\Lambda$.
We obtain a similar lower bound by taking only the term which is the closest to the supremum.
Hence, roughly speaking, for large $\Lambda$, the pressure $P_{\Lambda}$ is the Legendre transform $(F_{\Lambda})^*$ of the Helmholtz free energy, which becomes exact at infinite volume.
By taking another Legendre transform, we define the grand-canonical free energy as follows:
\begin{equation}\label{LT}
F^{\text{g.c.}}_{\Lambda}(\rho):=\sup_{z}\{\rho\ln z-P_{\Lambda}(z)\}.
\end{equation}
Note that 
if the thermodynamic limit of $F_{\Lambda}$ is convex, then by taking the Legendre transform twice we obtain:
$F^{\text{g.c.}}_\Lambda:=P_{\Lambda}^*=(F_{\Lambda})^{**}=F_{\Lambda}$, for large $\Lambda$, i.e., we recover the Helmholtz free energy and this is exact at infinite volume.
The supremum occurs at some $z^*$ such that $\rho=z^* P'_{\Lambda}(z^*)$ which
coincides with the definition of the thermodynamic density as in \eqref{density}.
In other words, given some value of the density $\rho$, there is an activity $z^*(\rho)$ that can produce it.
Using the inversion \eqref{invfor}, we also have a formula for it: $z^*(\rho)=\rho \e^{-\mathcal B'(\rho)}$.
Substituting back to \eqref{LT} we obtain:
\begin{eqnarray}\label{sMT}
F^{\text{g.c.}}_{\Lambda}(\rho) & = & \rho\ln(\rho \e^{-\mathcal B'(\rho)})-P_{\Lambda}(z^*)\nonumber\\
& = & \rho\ln\rho-\rho \mathcal B'(\rho)-P_{\Lambda}(z^*),
\end{eqnarray}
where $z^*=z^*(\rho)$.
Vice versa, we can also view $\rho$ as being dependent on the $z^*$ that produces it
and in particular recall the following expressions in terms of combinatorial species: that is
$\rho=\mathcal C^{\bullet}(z^*)$ and $P_{\Lambda}(z^*)=\mathcal C(z^*)$.
Next we use a formula known as the Dissymmetry Theorem, see \cite{bergeron-labelle-leroux1998book}, Section 4.2, Theorem 3.1:
\begin{equation}\label{dis}
\mathcal C^{\bullet}(z) + \mathcal B(\mathcal C^{\bullet}(z) )  =  \mathcal B^{\bullet}(\mathcal C^{\bullet}(z) )+\mathcal C(z),
\end{equation}
which evaluated at $z^*$ it gives:
\begin{equation}\label{dis2}
\rho + \mathcal B(\rho)  =  \rho \mathcal B'(\rho)+P_{\Lambda}(z^*).
\end{equation}
The proof is quite simple (see also \cite{jttu2014}, Theorem 3.3): fixing a connected graph one counts
how many times it occurs in each one of the four above combinatorial classes and finds that \eqref{dis2} is satisfied.
Substituting \eqref{dis2} into \eqref{sMT}
we obtain:
\begin{eqnarray}\label{sMT2}
F^{\text{g.c.}}_{\Lambda}(\rho) =
\rho\ln\rho-\rho - \mathcal B(\rho).
\end{eqnarray}
Note that this is true even at finite volume. In the infinite volume limit, since $F^{\text{g.c.}}=F^{**}=F$,
which is the infinite volume limit of the Helmholtz free energy, the above formula
holds for the latter as well, when it is strictly convex.

In a similar fashion we obtain the equation of state (pressure vs density):
given the free energy computed above in \eqref{sMT2} we take again the Legendre transform:
\begin{equation}\label{secondLT}
(F^{\text{g.c.}}_{\Lambda})^*(z) =  \sup_{\rho}\{ \rho\ln z-F^{\text{g.c.}}_{\Lambda}(\rho)\}.
\end{equation}
The supremum occurs at some $\rho^*$ such that $\ln z = F'_{\Lambda}(\rho^*)=\ln\rho^* - \mathcal B'(\rho^*)$ (by \eqref{sMT2}).
Notice also that this is \eqref{invfor}.
Substituting back to \eqref{secondLT} and using \eqref{sMT2} computed at $\rho^*$ we obtain:
\begin{eqnarray}\label{eos}
F^*_{\Lambda}(z) & = & \rho^* F'_{\Lambda}(\rho^*)-F_{\Lambda}(\rho^*) \\
& = & \rho^* -\rho^* \mathcal B'(\rho^*) + \mathcal B(\rho^*)\nonumber\\
& = & \rho^*-\sum_{n\geq 2} \frac{(n-1)}{n!}\beta_n(\rho^*)^n.
\end{eqnarray}
Now, if we start with $P_{\Lambda}$ being convex, then $(F^{\text{g.c.}}_{\Lambda})^*\equiv P_{\Lambda}$ and hence we obtain the equation of state:
\begin{equation}\label{finaleos}
\bar P_{\Lambda}(\rho):= P(z^*)=\rho -\rho \mathcal B'(\rho) - \mathcal B(\rho),
\end{equation}
which is true also at finite volume.
In both \eqref{sMT2} and \eqref{finaleos} the pending question is about the convergence of
the power series with coefficients the 2-connected graphs. A direct proof
based on an equivalent ``tree-graph" inequality \eqref{tgi} is missing. However,
indirect proofs have been established by using the inversion and then the composition of two power series, hence inheriting the radius of convergence of the inversion (as in Section~\ref{sec:inv}). 
The question is whether we can hope
for improvements by a more direct method.

In Example~\ref{ex1} we had a better convergence for $\bar P_{\Lambda}(\rho)$ than for
$P_{\Lambda}(z)$. Is this the case also here? The answer is negative, since in \eqref{invfor}
we assumed the convergence of $T(\rho)$ which brings in a similar constraint as for $P(z)$.
Then the question can be rephrased as whether there is a direct computation of $P(\rho)$ or $F_{\Lambda}(\rho)$ avoiding the ``problematic" inversion \eqref{invfor}.
A natural candidate is to work in the canonical ensemble for fixed density $\rho:=\frac{N}{|\Lambda|}$. This is possible and it will be discussed in the next section, but unfortunately there will still be a similar constraint for the convergence. The desired improvement seems to be possible
so far only in simple cases which are amenable to explicit calculations without need of expansions.
Nevertheless, it is still worth investigating the canonical ensemble being a direct method for computing free energies as well as helpful into elucidating the combinatorial structure of these expansions.

\begin{remark}
It is instructive to investigate the specific form of the Dissymmetry Theorem
in particular examples. One can consider the Tonks gas (hard rods in $d=1$), the multi-species Tonks gas \cite{jansen2015tonks},
the infinite dimensional gas ($d\to\infty$) \cite{FRW}, as well as the two-species hard spheres of small (microscopic) and big (macroscopic) size.
For the latter we refer to \cite{JT} for the expansion in the grand-canonical ensemble
and to \cite{NST} for the canonical ensemble.
\end{remark}

\begin{remark} One could use the results of this section
in order to represent the large deviations cost functional as a power series (in the regime 
when the latter convergences).
We start by computing the log moment generating function which corresponds to
an augmented grand-canonical partition function. Its logarithm
is the excess (due to the extra introduced activity) pressure of the system.
Then the large deviations rate functional is computed as the Legendre transform
and corresponds to the excess free energy. Cluster expansions would provide 
a power series expression of the above quantities, but only in the convergence regime. 
The same power series expansions can also be used to compute moderate deviations, see \cite{scola} and \cite{DS} for a general treatise.
\end{remark}

\section{Canonical ensemble}\label{sec:canonical}

We consider the canonical partition function \eqref{can_part_fcn}.
At first glance, the missing sum over $N$ seems obstructive in order to apply the previous combinatorial operations. However, this will be indeed possible but only after considering
new ``species". We introduce the set $\mathcal V:=\{V:\,V\subset [N]\}$ and, recalling the definition of the coefficients \eqref{phi} and \eqref{phit}, we choose $\mathbb X=\mathcal V$
and
\begin{equation}\label{choice}
h(V_i,V_j):=-\mathbf 1_{V_i\cap V_j\neq\emptyset},
\end{equation}
which yields:
\begin{equation}\label{phi2}
\phi(V_1,\ldots,V_n):=\prod_{1\leq i < j\leq n}\mathbf 1_{V_i\cap V_j=\emptyset}
\end{equation}
and 
\begin{equation}\label{phit2}
\phi^\mathsf T(V_1,\ldots,V_n):=\sum_{g\in\mathcal C_n}\prod_{\{i,j\}\in E(g)}(-\mathbf 1_{V_i\cap V_j\neq\emptyset}).
\end{equation}
Then we have:
\begin{eqnarray}\label{can}
Z_{\Lambda, N} & = & \frac{|\Lambda|^N}{N!}\int_{\Lambda^N} \sum_{g\in\mathcal G_N}
\prod_{\{i,j\}\in E(g)}f(x_i,x_j)\prod_{i=1}^N\frac{\dd x_i}{|\Lambda|}\nonumber\\
& = &
\frac{|\Lambda|^N}{N!}\int_{\Lambda^N}
\sum_{\{V_1,\ldots,V_n\}\in\Pi(1,\ldots,N)}\prod_{i=1}^n \left(\sum_{g \in \mathcal C_{V_i}}\prod_{\{i,j\}\in E(g)}f(x_i,x_j)\right)
\prod_{i=1}^N\frac{\dd x_i}{|\Lambda|}\nonumber\\
& = &
\frac{|\Lambda|^N}{N!}
\sum_{\substack{V_1,\ldots,V_n \\ V_i\cap V_j =\emptyset,\, \forall i\neq j}}\prod_{i=1}^n \zeta_{\Lambda}(V_i)\nonumber\\
& = & 
\frac{|\Lambda|^N}{N!}
\sum_{n=0}^{\infty} \frac{1}{n!}\sum_{(V_1,\ldots,V_n)}\phi(V_1,\ldots,V_n)
\prod_{i=1}^n \zeta_{\Lambda}(V_i)\nonumber\\
& = & 
\frac{|\Lambda|^N}{N!}
\exp\left\{
\sum_{n=1}^{\infty} \frac{1}{n!}\sum_{(V_1,\ldots,V_n)}\phi^\mathsf T(V_1,\ldots,V_n)
\prod_{i=1}^n \zeta_{\Lambda}(V_i)
\right\},
\end{eqnarray}
where
\begin{equation}\label{activity}
\zeta_{\Lambda}(V):= \sum_{g \in \mathcal C_{V}}\bar w_{\Lambda}(g),\qquad
\bar w_{\Lambda}(g):=\int_{\Lambda^{|V|}}\prod_{\{i,j\}\in E(g)}f(x_i,x_j)\prod_{i\in V} \frac{\dd x_i}{|\Lambda|}.
\end{equation}
Note that we introduced the new notation $\bar w_{\Lambda}$ which compared to \eqref{weight} has an integration over the normalized measure $\frac{dx}{|\Lambda|}$, i.e., $\bar w_{\Lambda}(g)=\frac{1}{|\Lambda|^{|g|}} w_{\Lambda}(g)$.
The key step is from line two to line three where we remove the constraint that the collection
$\{V_1,\ldots,V_n\}$ has to be a partition, thanks to the normalized measure, i.e., $\zeta_{\Lambda}(V)=1$, if $|V|=1$.
In this manner, in line four, we managed to express the (canonical) partition function as a ``grand-canonical" partition function, but for clusters $V\subset [N]$
with hard-core interaction and activity $\zeta_{\Lambda}(V)$.
Thus, from now on we will use the following generic formulation of the partition function to which we alluded in the beginning of the paper:
let $(\mathbb X, \mathcal X)$ be a measurable space and $\mathfrak M_\C(\mathbb X,\mathcal X)$ the set of complex linear combinations of $\sigma$-finite non-negative measures on $(\mathbb X,\mathcal X)$. 
When there is no risk of confusion, we shall write $\mathfrak M_\mathbb C$ for short. 
We consider the following generic partition function:
\begin{eqnarray}\label{genPT}
	\Xi_{\mathbb X}(z) & := & 1+\sum_{n=1}^\infty \frac{1}{n!}\int_{\mathbb X^n} \e^{-\beta  H_n(\vect x)} z^n(\dd \vect x)\nonumber\\
	& = & 1+\sum_{n=1}^\infty \frac{1}{n!}\int_{\mathbb X^n}\phi(x_1,\ldots,x_n)z^n(\dd \vect x),
\end{eqnarray}
where $\phi$ is given in \eqref{phi} and the measure $z$ can be thought of as adding an external potential $V_{\text{ext}}:\mathbb X\to\mathbb R$, i.e., as $z(\dd x)=\e^{-V_{\text{ext}}(x)}\dd x$.
For the above to be finite,
let us introduce an extra assumption on  $z \in \mathfrak M_\C(\mathbb X,\mathcal X)$, namely
\be \label{eq:fivo}
	\int_\mathbb X \e^{\beta B(x)}| z|(\dd x) < \infty.
\ee
This approach is followed in \cite{poghosyan-ueltschi2009}. Note that the grand-canonical partition function corresponds to the choice $\mathbb X=\Lambda\subset\mathbb R^d$ and $z(\dd x)=z \dd x$, while the canonical to the choice $\mathbb X=\mathcal V$ and a discrete measure $\zeta_{\Lambda}$ and it can also be seen as a direct application of the ``Abstract Polymer Model", \cite{KP}.
This more general set-up will also be useful for treating correlation functions.

The theorem proved in \cite{pulvirenti-tsagkaro2012} states that in computing the right hand side of \eqref{genPT} for the above choice of the canonical ensemble, there are several cancellations that lead us to $\mathcal B(\rho)$ plus some lower order terms in $|\Lambda|$ which vanish in the thermodynamic limit.
However, as far as convergence is concerned, we get again a similar radius of convergence
as in the grand-canonical approach despite the fact that this is a direct method.
This is due to the fact that we have power series with coefficients being expressed via connected graphs and again we need to use the tree-graph inequality in order to
prove convergence.
More precisely, by using \eqref{tgi} we have the following bound on the activity \eqref{activity}:
\begin{equation}\label{boa}
|\zeta_{\Lambda}(V)|\leq \e^{n B}n^{n-2}\frac{1}{|\Lambda|^{n-1}}C^{n-1}, \quad C:=\int_{\mathbb R^d}|f(0,q)|\dd q,
\end{equation}
for $|V|=n$ and where we used the simpler stability condition $\sum_{1\leq i<j\leq n} V(x_i, x_j)\geq -Bn$, instead of \eqref{stability}.
Then, the corresponding convergence condition \eqref{cond} for the general formulation \eqref{genPT} in the case of the canonical ensemble (by choosing $a(V)=c |V|$, for some $c>0$) reads:
\begin{eqnarray}\label{conv}
\sum_{V:\, V'\cap V\neq\emptyset}|\zeta_{\Lambda}(V')| \e^{c |V'|} & \leq &
|V| \sum_{n\geq 1}\e^{n B}\binom{N-1}{n-1} n^{n-2}\frac{1}{|\Lambda|^{n-1}}C^{n-1}\e^{cn}\nonumber\\
& \leq &
|V| \e^{c+B}\sum_{n\geq 1}\frac{n^{n-2}}{(n-1)!}\left(\frac{N-1}{|\Lambda|} \e^{c+B}C\right)^{n-1}
\leq c |V|,
\end{eqnarray}
where we have to choose $c>0$ accordingly and secure that
$\frac{N-1}{|\Lambda|} \e^{a+B}C<1$ so that the sum over $n$ is a convergent geometric series.
Hence,
the key idea is that one has to control the increasing (with $n$) combinatorics by reconstructing the density $\frac{N}{|\Lambda|}$.

On the other hand, the main idea for the cancellations leading to 2-connected graphs lies on the following re-writing of \eqref{can}:
\begin{eqnarray}
&& \frac{1}{|\Lambda|}\sum_{n=1}^{\infty} \frac{1}{n!}\sum_{\substack{(V_1,\ldots,V_n)\\ V_i\subset [N]}}\phi^\mathsf T(V_1,\ldots,V_n)
\prod_{i=1}^n \zeta_{\Lambda}(V_i)=\nonumber\\
&&
\frac{N}{|\Lambda|}\sum_{k\geq 1}
\frac{1}{k+1}\binom{N-1}{k}
\sum_{n=1}^{\infty} 
\frac{1}{n!}\sum_{\substack{(V_1,\ldots,V_n): \\ \cup_{i=1}^n V_i=[k+1]}}
\phi^\mathsf T(V_1,\ldots,V_n)
\prod_{i=1}^n \zeta_{\Lambda}(V_i),
\end{eqnarray}
where we chose $k+1$ labels among $N$, called them $[k+1]=\{1,\ldots,k+1\}$ and summed over all collections of polymers $V$ that span these labels.
We define:
\be\label{p9_1}
P_{N,|\Lambda|}(k):=\frac{(N-1)\ldots(N-k)}{|\Lambda|^k}
\ee
and
\be\label{p9_2}
B_{\beta,\Lambda}(k):=\frac{|\Lambda|^k}{k!}\sum_{n=1}^{\infty} 
\frac{1}{n!}\sum_{\substack{(V_1,\ldots,V_n): \\ \cup_{i=1}^n V_i=[k+1]}}
\phi^\mathsf T(V_1,\ldots,V_n)
\prod_{i=1}^n \zeta_{\Lambda}(V_i).
\ee
The key observation is that 
for all $k\geq 2$ we can split $B_{\beta,\Lambda}(k)$ as follows:
\be\label{p1.5}
B_{\beta,\Lambda}(k)= B^*_{\beta,\Lambda}(k)+ R_{\Lambda}(k),
\qquad B^*_{\beta,\Lambda}(k):=\frac{|\Lambda|^k}{k!}\sum_{n=1}^{\infty} 
\frac{1}{n!}\sum^*_{\substack{(V_1,\ldots,V_n): \\ \cup_{i=1}^n V_i=[k+1]}}
\phi^\mathsf T(V_1,\ldots,V_n)
\prod_{i=1}^n \zeta_{\Lambda}(V_i),
\ee
where  the $\sum^*$ is now a finite sum (as well as the sum over $n$) and
contains all ordered sequences $(V_1,\ldots,V_n)$ 
which satisfy the following properties: 
\begin{eqnarray}
&& V_i\neq V_j,\,\forall i\neq j\quad \text{and}\label{p1.3}\\
&& n+1= \sum_{i=1}^n (|V_i|-1)+1\label{p1.4}.
\end{eqnarray}
The remainder $R_{\Lambda}(k)$ vanishes in the thermodynamic limit as when we have many overlaps of labels we obtain more factors $\frac{1}{|\Lambda|}$ from the normalized measure.
For $B_{\beta,\Lambda}^*$ the following simplification is possible {\it at finite volume with periodic boundary conditions}: given a graph $g\in\mathcal C_{k+1}$
we define $\mathbb B(g):=\{b_1,\ldots,b_r\}$, where $b_i$, $i=1,\ldots,r$ are the
2-connected components (blocks) of $g$. We denote by $\mathcal F_{\neq}(g)$ the collection of 
all $F\subset\mathbb B(g)$ such that $\cup_{b\in F}b$ is a connected graph.
We also define $\mathcal H(g)$ to be the collection of all such graphs
\begin{equation}\label{H}
\mathcal H(g):=\{
g':g'=\cup_{b\in F}b, F\in\mathcal F_{\neq}(g)
\}
\end{equation}
and similarly,
\begin{equation}\label{calA}
\mathcal A(g):=\{
V(g'):g'\in\mathcal H(g)
\}.
\end{equation}
With this definition we obtain:
\be\label{may14}
B^*_{\beta,\Lambda}(k) =\frac{|\Lambda|^k}{k!}\sum_{g\in\mathcal C_{k+1}}\bar w_{\Lambda}(g)
\sum^*_{\substack{(V_1,\ldots,V_n): \, V_i\in \mathcal A(g) \\
\cup_{i=1}^n V_i=[k+1]
}}\phi^\mathsf T(V_1,\ldots,V_n)
= \frac{|\Lambda|^k}{k!}\sum_{g\in\mathcal B_{k+1}}\bar w_{\Lambda}(g),
\ee
under periodic boundary conditions in the integration over $\Lambda$
in $\bar w_{\Lambda}$.
For the proof we refer to \cite{pulvirenti-tsagkaro2012}, Section 5.
Note that this reduction to more connected structures will be observed
also in the case of correlation functions in the canonical ensemble as in the next section.
We summarize these facts in the following theorem:

\begin{theorem}\label{thm1}
There exists a constant $c_0\equiv c_0(\beta,B)>0$ independent of $N$ and $\Lambda$ 
(see also Remark~\ref{R1}) such that if $\rho\,C(\beta)<c_0$ 
then for the canonical partition function \eqref{can_part_fcn} with periodic boundary conditions we have:
\be\label{p3.1}
\frac{1}{|\Lambda|}\log Z^{per}_{\beta,\Lambda,N}=
\log\frac{|\Lambda|^N}{N!}+
\frac{N}{|\Lambda|}\sum_{k\geq 1} \frac{1}{k+1}P_{N,|\Lambda|}(k)B_{\beta,\Lambda}(k),
\ee
as given in \eqref{p9_1}, \eqref{p9_2} and
with $N=\lfloor\rho|\Lambda|\rfloor$. In the thermodynamic limit we obtain:
\be\label{n3}
\lim_{N,|\Lambda|\to\infty,\,N=\lfloor\rho|\Lambda|\rfloor}P_{N,|\Lambda|}(k)B_{\beta,\Lambda}(k)=
\rho^{k}\beta_k,
\ee
for all $k\geq 1$
and $\beta_k$ given in \eqref{betan}.
Furthermore, there exist constants $C, c>0$ such that, for every $N$ and $\Lambda$,
and $k\geq 1$ we have: 
\be\label{n1}
\frac{1}{k+1}P_{N,|\Lambda|}(k)|B_{\beta,\Lambda}(k)|\leq C e^{-ck}
\ee
and we can exchange the limit and the sum in \eqref{p3.1}.
\end{theorem}

\begin{remark}\label{R1}
Using the more recent tree-graph inequality \cite{procacci-yuhjtman2017} as well as other results \cite{NF20, FP, fernandez-procacci-scoppola2007, MProc} one can improve the value of $c_0$ obtained in the original \cite{pulvirenti-tsagkaro2012} paper, but it will only
be a minor improvement and
it will not avoid the singularity present in the expansion of the
pressure with respect to the activity despite the fact that it is a direct method.
It might be worth investigating how one could use the perturbative method of the cluster expansion around a different 
point than the ideal gas, as it was also hinted in \cite{MP22} where the authors could obtain
some significant improvement for the analyticity of the pressure (but not for the cluster expansion).
\end{remark}

In the rest of the paper, using the generic form \eqref{genPT} of the partition function
we will study expansions of the correlation functions
as they appear in the literature of the liquid state theory. We will prove convergence of the corresponding power series expansions in both
the density and the activity.
We stress again that this is valid always in 
a small region around the ideal gas, i.e., in the gas phase, even though in practice
some of them are used in denser regimes. We will conclude with a quick discussion
on the rigorous justification of the closures.

\section{Correlation functions}\label{sec:correlations}

As an application of the more general (functional analytic) framework, we consider expansions of various correlation functions in terms of the activity as well as the density. Special interest will be given
to the case of two-point correlation function and in particular to the direct correlation function.
We also compare with the expansion in the canonical ensemble which
exhibits a similar structure as for the free energy. We conclude with a discussion about
closures.
One of the goals is to provide the mathematical tools that can turn ``graphical descriptions" into rigorous proofs about the converge of the corresponding series. We start with the one discussed in Section~\ref{sec:inv}.

\subsection{One-point correlation function in the grand-canonical ensemble}\label{sec:one_point}

The one-particle density viewed as a measure is given by:
\be \label{eq_density}
	\rho(\dd q;z):= \frac{1}{\Xi_{\mathbb X}(\beta,z)} \Biggl(1+ \sum_{n=1}^\infty \frac{1}{n!} \int_{\mathbb X^n}\e^{- \beta H_{n+1}(q,x_1,\ldots, x_n)} z^n(\dd \vect x) \Biggr) z(\dd q). 
\ee
Notice that thanks to the more general formulation of the partition function as in \eqref{genPT}
it can also be expressed as
\be\label{eq:difflog}
	\rho(\dd q;z) =\Bigl( \frac{\delta}{\delta z(q)} \log \Xi(\beta, z)\Bigr) z(\dd q),
\ee
in analogy to \eqref{density}.
Similarly to \eqref{inv} we also have that
\be \label{eq:rhoa}
 	\rho[z](dq)\equiv\rho(\dd q;z) = \e^{- A(q;z)} z(\dd q),
 \ee
where $A$ is given in \eqref{A}.
Note that we use the notation $z\mapsto\rho[z]$ when we want to view it as a map $\rho:\mathfrak M_\C\to\mathfrak M_\C$.
Note that the convergence of $A$ is similar to
Proposition~\ref{propM1} under a bit more restrictive condition \eqref{PU} 
that also controls the first factor of \eqref{A1}. 
This is summarized in the next lemma:
	
\begin{lemma} \label{lem:prelim}
	Let $A_n(q;x_1,\ldots,x_n)$ be as in~\eqref{A1} and define $A(q;z)$ as in~\eqref{A}. Let $z\in \mathfrak M_\C$ satisfy only
	\be \label{PU}
		\int_\mathbb X \bar f(x,y)\, \e^{a(y)+ \beta B(y)}|z|(\dd y)\leq a(x),
	\ee
	for some weight function $a:\mathbb X\to \R_+$ and all $x\in \mathbb X$. Then $z$ is in the domain of convergence $\mathscr D(A)$. 
	
	If in addition $z$ satisfies the finite-volume condition~\eqref{eq:fivo}, then the density $\rho(\dd q;z)$ defined in~\eqref{eq_density} is equal to $\exp( - A(q;z)) z(\dd q)$, moreover
	\begin{align}
		\log \Xi_{\mathbb X}(\beta,z) & = \sum_{n=1}^\infty \frac{1}{n!}\int_{\mathbb X^n}  \phi^\mathsf T(x_1,\ldots,x_n) z^n(\dd \vect x),\label{lnXi}\\
		\rho(\dd q;z) & = z(\dd q)\Biggl( 1+ \sum_{n=1}^\infty \frac{1}{n!}\int_{\mathbb X^n} \phi^\mathsf T(q,x_1,\ldots,x_n) z^n(\dd \vect x)\Biggr),\label{rho}
	\end{align} 
	with absolutely convergent integrals and series.
\end{lemma}

\noindent The ingredients of the proof are again the tree-graph inequality due to~\cite{procacci-yuhjtman2017} and the fact that fixing the origin we partition over the remaining labels (for a similar implementation see~\cite[Eq. (4.17)]{jttu2014}).
See also \cite{JK} for a more general convergence condition for the activity expansion of correlation functions.

Similarly to Section~\ref{sec:inv}
we can invert the map $z\mapsto \rho[z]$ and express the inverse with 2-connected graphs.  
Here, in addition to \eqref{stability}, we also assume that for all $x\in \mathbb X$  and some function $B^*:\mathbb X\to \R_+$ we have
\begin{equation}\label{stabextra}
	\inf_{y\in \mathbb X} V(x,y)\geq - B^*(x).
\end{equation}
Then, getting prepared for the rigorous statement of the inversion we first introduce
the following condition on $\nu\in \mathfrak M_\C$:
\be \label{suffsuff}
		\int_\mathbb X \bar f(x,y)\, \e^{a(y)+b(y)+ \beta B(y)+\beta B^*(y)}|\nu|(\dd y)\leq a(x).
	\ee
Comparing to \eqref{PU} one might question the extra terms in the exponent. The answer is
that convergence condition for the inverted power series requires not only the validity of the resulting series but also the fact that taking again its inverse and getting back to the original series the resulting value is within the radius of convergence of the latter.
For completeness we have decided to present the exact conditions,
even though we are aware that they can be appreciated only when one dives into the proofs, which can be found in
the original reference \cite{JKT_JFA}.
Define $\mathsf V_b$ by
$$
	\mathsf V_b = \bigl\{ \nu \in \mathfrak M_\C\mid \exists a:\mathbb X\to \R_+:\, a\leq b,\ \nu\text{ satisfies~\eqref{suffsuff}}\bigr\}. 
$$
The rigorous statement for the inversion discussed in \eqref{invfor} is given in the following theorem:
\begin{theorem} \label{thm:virmain2}
	There is a set $\mathsf U_b\subset \mathscr D(A) \subset \mathfrak M_\C$ such that $z\mapsto \rho[z]$ is a bijection from $\mathsf U_b$ onto $\mathsf V_b$, and for every $z\in \mathsf U_b$, $\nu\in \mathsf V_b$, we have $\rho[z]=\nu$ if and only if 
	\be \label{eq:virmain2a} 
		z(\dd q ) = \nu(\dd q) \exp\Biggl(-  \sum_{n=1}^\infty \frac{1}{n!}\int_{\mathbb X^n}  D_{n+1}(q,x_1,\ldots, x_n) \nu(\dd x_1) \cdots \nu(\dd x_n) \Biggr),
	\ee
where the latter converges.

If $z\in \mathfrak M_\C$ fulfills \eqref{suffsuff} for all $x\in \mathbb X$ and
for some $a\leq b$ with $\e^{a}|z|\in \mathsf V_b$ for the same functions $a$ and $b$, 
then  $\rho[z]\in \mathsf V_b$ and hence $z \in \mathsf U_b$. 

If instead the following conditions including also a ``finite volume condition" hold, 
	\be \label{dissymmetry-condition:b}
		\int_{\mathbb X}\bar f(x,y)\, \e^{a(y)+ \beta B(y)} |z|(\dd y) \leq a(x),\quad 	\e^{a+\beta B}|z|\in \mathsf V_b, \quad \int_{\mathbb X}(1+ b(q)) \e^{a(q)+ \beta B(q)}|z|(\dd q)<\infty,
	\ee
	then also
	\be \label{eq:virmain2b}
		\log \Xi_{\mathbb X}(\beta,z) = \int_\mathbb X\rho(\dd x_1;z) 
			- \sum_{n=2}^\infty \frac{1}{n!}\int_{\mathbb X^n} (n-1) D_n(x_1,\ldots,x_n) \prod_{i=1}^n \rho(\dd x_i;z).
	\ee
\end{theorem}

\noindent
This theorem is proved in \cite{JKT_JFA} by using the construction described in Section~\ref{sec:inv}.
We obtain that \eqref{tree-eq} is the solution of the fixed point equation \eqref{FP} which proves \eqref{eq:virmain2a}. The rest is to verify that computing the density from the activity and back produces bounded quantities in the relevant spaces.
Last, \eqref{eq:virmain2b} is obtained from the dissymmetry theorem presented in Section~\ref{sec:M2} 
again by assuming some extra conditions \eqref{dissymmetry-condition:b}
that guarantee that all involved quantities are bounded.

\begin{remark}
Formula \eqref{eq:virmain2b} does not make any sense in the ``infinite volume case" even if we consider the translation invariant case as discussed below \eqref{eq:fivo}. In this case, though, the right hand side is proportional to the  volume of $\mathbb{X}$, up to boundary errors. Hence, $\log \Xi(\beta,z)$ divided by the volume has a well defined limit. 
\end{remark}

\begin{remark}
In the sequel we will use the following terminology: we will say that
the operation $\frac{\delta}{\delta z(x_i)}$ in \eqref{eq:rhoa} makes the vertex $x_i$ ``white" and it corresponds to the vertex with variable $q$ in \eqref{eq:virmain2a}.
\end{remark}

Furthermore, similarly to Section~\ref{sec:M2} we can
define and compute the grand-canonical free energy.
The proofs follow the same reasoning as in the scalar case. 
We fix a reference measure $m(\dd x)$ on $\mathbb X$ (for example, the Lebesgue measure on $\R^d$). The (grand-canonical) free energy $\mathcal F_{\mathrm{GC}}[\nu]$ of a given density profile $\nu \in \mathfrak M_\C$ is defined via the Legendre transform of $\log \Xi(z)$ as 
\begin{equation}\label{GCFELT}
	\beta \mathcal F_{\mathrm{GC}}[\nu]:= \sup_{z}\Bigl(\int_\mathbb X \log \frac{\dd z}{\dd m}(x) \nu(\dd x) - \log \Xi(z)\Bigr),
\end{equation}
with $\frac{\dd z}{\dd m}$ the Radon-Nikod{\'y}m derivative of $z$ with respect to the reference measure $m$. The supremum in~\eqref{GCFELT} is over all non-negative measures $z\in \mathfrak M_\C$ that are absolutely continuous with respect to $m$ and such that the integral with the logarithm is absolutely convergent. 

\begin{theorem}\label{thm:GCFE}
	Assume that $\nu \in \mathsf V_b\cap \mathfrak M_\C$ is absolutely continuous with respect to $m$ and satisfies 
	\be \label{eq:nuassumption}
		\int_{\mathbb X}(1+b(q)) \nu(\dd q)<\infty,\quad 
		\int_\mathbb X\Bigl| \log \frac{\dd \nu}{\dd m}\Bigr|\dd \nu <\infty,\quad \int_\mathbb X \e^{\beta B+ b} \dd \nu <\infty, 
	\ee
	then 
	\begin{equation} \label{eq:free-energy}
		\beta \mathcal F_{\mathrm{GC}}[\nu]
			 = \int_\mathbb X   \bigl[\log \frac{\dd \nu}{\dd m}(x) - 1\bigr]\nu(\dd x)  - \sum_{n=2}^\infty \frac{1}{n!} \int_{\mathbb X^n}D_n\bigl(x_1,\ldots, x_n\bigr) \nu^n(\dd \vect x)
	\end{equation} 
	with absolutely convergent integrals and sum. 
\end{theorem} 

For the proof, we refer again to \cite{JKT_JFA}, Theorem 3.6.
Our goal in the following sections
is to use this inversion method in order to rigorously prove the validity
of other commonly used expansions in terms of the density. Alternatively, one could work directly in the canonical ensemble as illustrated in Section~\ref{sec:canonical} and we show how this can be adapted to the correlation functions. We start with the $n$-point correlation function.

\subsection{General $n$-point correlation function}

Similarly to \eqref{eq_density} we have:
\begin{equation}\label{twopointgen}
\rho^{(n)}_{\mathbb X}(\dd x_1,\ldots, \dd x_n; z):=
\frac{\prod_{i=1}^n z(\dd x_i)}{\Xi_{\mathbb X}(\beta, z)}
\sum_{N=n}^{\infty}\frac{1}{(N-n)!}\int_{\Lambda^{N-n}}  \e^{-\beta H_{N}(\mathbf x)}z(\dd x_{n+1})\ldots z(\dd x_N).
\end{equation}
For simplicity, we can also 
view it as
the absolutely continuous part with respect to the Lebesgue measure of the previous measure formulation.
The following formula can be verified by functional differentiation:
\begin{equation}\label{rhon_gen}
\rho^{(n)}_{\mathbb X}(x_1,\ldots, x_n; z)=\frac{1}{\Xi_{\mathbb X}(\beta, z)}\frac{\delta^n \Xi_{\mathbb X}(\beta, z)}{\delta z(x_1)\ldots\delta z(x_n)}
\prod_{i=1}^n z(x_i).
\end{equation}
Note that this is compatible with \eqref{eq:difflog} (considering the absolutely continuous
with respect to Lebesgue part of the measures) for $n=1$:
\begin{equation}\label{rhon1}
\rho^{(1)}_{\mathbb X}(x_1; z)=\frac{1}{\Xi_{\mathbb X}(\beta, z)}\frac{\delta \Xi_{\mathbb X}(\beta, z)}{\delta z(x_1)}
z(x_1)=\frac{\delta \ln\Xi_{\mathbb X}(\beta, z)}{\delta z(x_1)}z(x_1).
\end{equation}
For higher $n$ the structure is different, e.g. for $n=2$ we have:
\begin{equation}\label{diff}
\frac{1}{\Xi_{\mathbb X}(\beta, z)}\frac{\delta^2 \Xi_{\mathbb X}(\beta, z)}{\delta z(x_1)\delta z(x_2)}=\frac{\delta}{\delta z(x_1)}\frac{\delta \ln\Xi_{\mathbb X}(\beta, z)}{\delta z(x_2)}
+\frac{1}{\Xi_{\mathbb X}(\beta, z)}\frac{\delta \Xi_{\mathbb X}(\beta, z)}{\delta z(x_1)}\frac{1}{\Xi_{\mathbb X}(\beta, z)}\frac{\delta \Xi_{\mathbb X}(\beta, z)}{\delta z(x_2)}.
\end{equation}
Hence, it is useful to introduce the following quantity:
\begin{equation}\label{un}
u^{(n)}_{\mathbb X}(x_1,\ldots,x_n;z):=\frac{\delta^n \ln\Xi_{\mathbb X}(\beta, z)}{\delta z(x_1)\ldots\delta z(x_n)}\prod_{i=1}^n z(x_i).
\end{equation}
In fact, the correlation functions $\rho^{(n)}_{\mathbb X}$ and $u^{(n)}_{\mathbb X}$ can
be viewed as the Taylor coefficients for the expansions around $0$ of $\Xi_{\mathbb X}$ and $\ln\Xi_{\mathbb X}$, respectively.
Hence they should also be related as follows:
\begin{equation}\label{rhoandu}
\rho^{(n)}_{\mathbb X}(x_1,\ldots,x_n;z)=\sum_{k=1}^n\sum_{\substack{\{P_1,\ldots,P_k\} \\\in\Pi(1,\ldots,n)}}\prod_{i=1}^k u^{(|P_i|)}_{\mathbb X}(\mathbf x_{P_i};z),
\end{equation}
where we recall that $\Pi_k(1,\ldots,n)$ is the set of all partitions of $\{1,\ldots,n\}$ into $k$ blocks. 
Indeed, given $\rho^{(n)}_{\mathbb X}$ defined in \eqref{rhon_gen}, formula \eqref{rhoandu} is equivalent to \eqref{un} and the functions $u^{(n)}_{\mathbb X}$ are called truncated correlation functions (or cluster or Ursell functions).
For example, for $n=2$ from \eqref{diff} we have:
\begin{equation*}
u^{(2)}_{\mathbb X}(x_1,x_2)=\rho^{(2)}_{\mathbb X}(x_1,x_2)-\rho_{\mathbb X}^{(1)}(x_1)\rho_{\mathbb X}^{(1)}(x_2).
\end{equation*}
It will be proved useful to consider the following normalized versions:
\begin{equation}\label{g}
g^{(n)}_{\mathbb X}(x_1,\ldots,x_n):=\frac{\rho_{\mathbb X}^{(n)}(x_1,\ldots,x_n)}{\prod_{i=1}^n\rho_{\mathbb X}^{(1)}(x_i)},
\qquad n\geq 1,
\end{equation}
\begin{equation}\label{h}
h^{(n)}_{\mathbb X}(x_1,\ldots,x_n):=\frac{u^{(n)}_{\mathbb X}(x_1,\ldots,x_n)}{\prod_{i=1}^n\rho_{\mathbb X}^{(1)}(x_i)}, \qquad n\geq 2
\end{equation}
and
\begin{equation}\label{h1}
h^{(1)}_{\mathbb X}(q):=\ln\frac{u^{(1)}_{\mathbb X}(q)}{z(q)}.
\end{equation}

Recall that we represented the operation $\frac{\delta}{\delta z(x_1)}$ by choosing
the vertex $x_1$ and making it ``white". Then from \eqref{un}
the power series expansion in $z$ is a rather straightforward application of the operations $\frac{\delta}{\delta z(x_1)},\ldots,\frac{\delta}{\delta z(x_n)}$ to $\ln\Xi_{\mathbb X}$ (as 
given in \eqref{lnXi}) which yields:
\begin{equation}\label{psu}
u^{(n)}_{\mathbb X}(x_1,\ldots,x_n;z)=
\left(
1+ \sum_{k=1}^\infty \frac{1}{k!}\int_{\mathbb X^k} \phi^\mathsf T(x_1,\ldots,x_n,x_{n+1},\ldots, x_{n+k}) z^k(\dd \vect x)\right)\prod_{i=1}^n z(x_i).
\end{equation}
Note that in the particular case $\mathbb X=\Lambda$ and $z(\dd x)=z \dd x$ we obtain:
\begin{equation}\label{psu2}
u^{(n)}_{\Lambda}(x_1,\ldots,x_n;z)
=\frac{z^n}{n!}
\sum_{k\geq 0}\frac{z^k}{k!}\sum_{g\in\mathcal C_{n,n+k}} w^{\bullet}_{\Lambda}(g;x_1,\ldots , x_n),
\end{equation}
where
$\mathcal C_{n,n+k}$ is the set of connected graphs with $n$ white and $k$ black vertices
and
\be\label{act_with_points}
w^{\bullet}_{\Lambda}(g;x_1,\ldots , x_n):=
\int_{\Lambda^k}  
 \prod_{ \{i,j\} \in E(g)} f(x_i,x_j)
 \prod_{j=n+1}^{n+k} dx_j.
\ee 
Note the difference in the notation for $w_{\Lambda}$, $\bar w_{\Lambda}$ and $w_{\Lambda}^{\bullet}$:
we denote with a $^\bullet$ whenever some particles are fixed: $x_1,\ldots,x_n$.
The convergence can be proved in a similar way as for the case $n=1$ in Lemma~\ref{lem:prelim} by operating on the vertices $x_1,\ldots, x_n$, one at a time.

On the other hand, for the correlation functions in view of \eqref{rhoandu} we have:
\begin{equation}\label{rhon}
\rho^{(n)}_{\Lambda}(x_1,\ldots,x_n;z)=\frac{z^n}{n!}\sum_{k\geq 0}\frac{z^k}{k!}\sum_{g\in\mathcal G_{n,n+k}} w^{\bullet}_{\Lambda}(g;x_1,\ldots , x_n),
\end{equation}
where in the set $\mathcal G_{n,n+k}$ of simple graphs with $n$ white and $k$ black vertices, there is a path from each black vertex to a white one. 

The next step is by using the theory developed in the previous section to replace the $z$ vertices by $\rho$ vertices. This can be found in \cite{morita-hiroike3, stell1964} and here we wish to give a strategy how to prove convergence.
As in the case $n=1$ we look for the ``reduced" structure which includes the white vertices
and the parts that ``fall out" from a given root-vertex which, after removing the part that falls out, it will become a $\rho$ vertex. 
The ``reduced" structure is found as follows: given a black vertex we check whether there are two ``independent" paths to two different white vertices. If not, then starting from the black vertex, there is first a part with some common edges, after which the second vertex of the last common edge has the above property. Turning that last vertex into a $\rho$ vertex is equivalent to
removing the hanging part. Repeating this procedure we obtain graphs with the $n$ fixed white vertices
and some black vertices (say $k$ many) that have the previous property, namely that there are two independent paths to any two different white vertices: these are called ``articulation free" graphs and we denote them
with $\mathcal B^{\text{AF}}_{n,n+k}$.
We obtain:
\begin{equation}\label{hwithrho}
h^{(n)}_{\Lambda}(x_1,\ldots,x_n)=\frac{1}{n!}\sum_{k\geq 0}\frac{\rho^k}{k!}\sum_{g\in\mathcal B^{AF}_{n,n+k}} w^{\bullet}_{\Lambda}(g;x_1,\ldots , x_n).
\end{equation}
Note that the factor $\frac{1}{n!}$ amounts to the fact that in the sum over $g$ we can permute the indices of the white vertices. We are not aware of any direct proof of counting
the cardinality of the articulation free graphs, neither of any equivalent tree-graph inequality involving them,
so in order to prove convergence of \eqref{hwithrho}
we need to adapt the strategy in Section~\ref{sec: invtrees} to the case of $n$ vertices.
Taking $\frac{\delta}{\delta z(q_2)}$ in \eqref{stellandco} (with $q_1$ instead of $0$) 
and applying it to
\eqref{h} we obtain:
\begin{eqnarray}\label{derstellandco}
h^{(2)}_{\mathbb X}(q_1,q_2) & = & \frac{1}{\rho_{\mathbb X}^{(1)}(q_1)\rho_{\mathbb X}^{(1)}(q_2)}
\frac{\delta^2 \ln\Xi_{\mathbb X}(\beta, z)}{\delta z(q_1)\delta z(q_2)}z(q_1)z(q_2)
\nonumber\\
& = &
\frac{1}{\rho_{\mathbb X}^{(1)}(q_1)\rho_{\mathbb X}^{(1)}(q_2)}
\frac{\delta}{\delta z(q_2)}\e^{-A(q_1;z)}z(q_1)z(q_2)
=
-\frac{1}{\rho_{\mathbb X}^{(1)}(q_2)}
\frac{\delta}{\delta z(q_2)}A(q_1;z) z(q_2)\nonumber\\
& = &
\sum_{n=0}^\infty \frac{1}{n!}\int_{\mathbb X^n} D_{n+2}(q_1, q_2, x_1, \ldots, x_n)
\prod_{i=1}^n \e^{- A(x_i;z)}z^n(\dd \vect x)+\nonumber\\
&&
+\int_{\mathbb X}\left\{\sum_{n=0}^\infty \frac{1}{n!}\int_{\mathbb X^n} D_{n+2}(q_1, q_3, x_1,\ldots,x_n)\prod_{i=1}^n \e^{- A(x_i;z)}  z^n(\dd \vect x)\right\}
\times
\nonumber\\
&&
\frac{1}{\rho_{\mathbb X}^{(1)}(q_2)}\e^{- A(q_3;z)}z(q_3)
z(q_2)\frac{\delta}{\delta z(q_2)}A(q_3;z)\dd q_3.
\end{eqnarray}
Hence, the question is if to each of the $x_{j}$, $j=1,\ldots,n$ we apply the power series $\bar T^{\circ}_{x_{j}}(\rho)$ whether we obtain \eqref{hwithrho}.
We have two contributions: the first term in the last equality of \eqref{derstellandco} which
consists of 2-connected graphs
and the second which has an ``intermediate" point $q_3$ (which in Section~\ref{sec:direct} will be called nodal). We called this class ``articulation free".
For $n\geq 3$ we take more derivatives, pointing to other labels, and obtain similar contributions: either an overall 2-connected graph or articulation free graphs allowing for nodal points between any choice of two fixed labels. The convergence should be again the
result of composition between the new kernel corresponding to the left hand side of \eqref{derstellandco} and the trees $z=\rho_{\Lambda}\bar T^{\circ}_0(\rho_{\Lambda})$
as in \eqref{comp}.
We hope to address
all this in detail in a forthcoming work.

\subsection{General $n$-point correlation function in the canonical ensemble}

Whenever we want to obtain an expansion in terms of the density, instead of entering into a combinatorially involved inversion procedure as in the previous section,
we can investigate the alternative option of working directly in the canonical ensemble.
This was established for the case of the pressure in Section~\ref{sec:canonical} and we present it here for the case of the various correlation functions. 
Although some of the details of the proof are different, the main strategy remains the same.
For a complete presentation we refer to \cite{kuna-tsagkaro2016} (see also \cite{pulvirenti-tsagkaro2015} for an earlier attempt).

The $n$-point canonical correlation function (for a fixed number of $N$ particles) is given by:
\begin{equation}\label{twopoint}
\rho^{(n)}_{\Lambda,N}(x_1,\ldots,x_n):=\frac{1}{(N-n)!}\int_{\Lambda^{N-n}}  \frac{1}{Z_{\Lambda,N}}\e^{-\beta H_{N}(\mathbf x)}\dd x_{n+1}\ldots \dd x_N.
\end{equation}
Note that $\rho_{\Lambda,N}^{(0)}=1$ and $\rho^{(1)}_{\Lambda,N}= \frac{N}{|\Lambda|}$ (with periodic boundary conditions). Thus, in the thermodynamic limit of $\Lambda\to\mathbb R^d$, $N\to\infty$ such that $\frac{N}{|\Lambda|}\to\rho$, we obtain that $\rho^{(1)}=\rho$.
For comparison purposes, note that the corresponding $n$-point correlation function for the grand-canonical ensemble originally given in \eqref{twopointgen} in the case
 $\mathbb X=\Lambda$ and $z(\dd x)=z \dd x$ it
is related to \eqref{twopoint} by
\begin{equation}\label{twopointgc}
\rho^{(n)}_{\Lambda}(x_1,\ldots,x_n; z):=\sum_{N=n}^{\infty}z^N\frac{Z_{\Lambda,N}}{\Xi_{\Lambda, \beta}(z)}\rho^{(n)}_{\Lambda,N}(x_1,\ldots,x_n)\prod_{i=1}^n z(x_i).
\end{equation}
Similarly to \eqref{g} and \eqref{h}
we will see that in the thermodynamic limit the leading order of the functions $\rho^{(n)}$ and $u^{(n)}$ (limits of
$\rho^{(n)}_{\Lambda,N}$ and $u^{(n)}_{\Lambda,N}$) is $\rho^n$. 
Hence, it is a common practice to introduce the following order one functions:
\begin{equation}\label{gcan}
g^{(n)}_{\Lambda,N}(x_1,\ldots,x_n):=\frac{\rho_{\Lambda,N}^{(n)}(x_1,\ldots,x_n)}{\rho^n},
\qquad n\geq 1,
\end{equation}
and
\begin{equation}\label{hcan}
h^{(n)}_{\Lambda,N}(x_1,\ldots,x_n):=\frac{u^{(n)}_{\Lambda,N}(x_1,\ldots,x_n)}{\rho^n}, \qquad n\geq 2.
\end{equation}
Due to the periodic boundary conditions all correlation functions introduced above will be invariant under translation.

We consider the following extension
of the canonical partition function, sometimes called the Bogolyubov functional, see \cite{B}, equation (2.11):
\begin{equation}\label{bog}
L_N(\phi) :=\frac{1}{Z_{\Lambda,N,\beta}}\frac{1}{N!}\int_{\Lambda^N}  \prod_{k=1}^N (1+\phi(x_k))\e^{-\beta H_N(\mathbf x)}\dd \mathbf x.
\end{equation}
Note that equivalently to \eqref{rhon} and \eqref{un}, where $\rho^{(n)}_{\Lambda}(x_1,\ldots,x_n;z)$ and $u^{(n)}_{\Lambda}(x_1,\ldots,x_n;z)$ are the Taylor expansion coefficients (in terms of formal power series and variational derivatives) of $\Xi_{\Lambda,\beta}$ and $\ln\Xi_{\Lambda,\beta}$, respectively, a similar property is also true for their ``canonical" version. In the latter case, the correlations $\rho^{(n)}_{\Lambda,N}(x_1,\ldots,x_n)$ and the truncated correlations $u^{(n)}_{\Lambda,N}(x_1,\ldots,x_n)$ are the coefficients of the following power series expressions of $L_N(\phi)$ and $\ln L_N(\phi)$:
\begin{equation}\label{bogphi}
L_N(\phi) = 1+\sum_{n=1}^{N}\frac{1}{n!}\int_{\Lambda^n}\phi(x_1)\ldots \phi(x_n)
\rho^{(n)}_{\Lambda,N}(x_1,\ldots,x_n)\dd\mathbf x
\end{equation}
and
\begin{equation}\label{lnbogphi}
\ln L_N(\phi) = \sum_{n=1}^{\infty}\frac{1}{n!}\int_{\Lambda^n}\phi(x_1)\ldots \phi(x_n)
u^{(n)}_{\Lambda,N}(x_1,\ldots,x_n)\dd\mathbf x.
\end{equation}
Note that while the first is in agreement with \eqref{twopoint} and \eqref{bog}, the second
can be viewed as the definition of $u^{(n)}_{\Lambda,N}$ which is also in agreement
with the corresponding relation \eqref{rhoandu}.
Having established this connection, we can compute $u^{(n)}_{\Lambda,N}$ and subsequently $h^{(n)}_{\Lambda,N}$ by applying the strategy of Section~\ref{sec:canonical}
to the ``augmented" partition function \eqref{bog}.
Then, by comparing with \eqref{lnbogphi} we obtain
the following theorem whose proof is given in \cite{kuna-tsagkaro2016}:

\begin{theorem}\label{thm2}
There exists a constant $c_0>0$ such that for all $\rho\,C(\beta)<c_0$ 
we have:
\begin{equation}\label{Stell_hn}
h^{(n)}(x_{1},\ldots,x_{n}):=\lim_{\substack{\Lambda\uparrow\mathbb R^{d},N\to\infty, \\ N=\lfloor\rho|\Lambda|\rfloor}}
h^{(n)}_{\Lambda, N}(x_{1},\ldots,x_{n})
=\sum_{k\geq 0} \rho^k \frac{1}{n!k!} \sum_{g \in \mathcal B^{\text{AF}}_{n,n+k}}
w^{\bullet}(g;q_1,\ldots , q_n),
\end{equation}
where 
\begin{equation}\label{act_with_points_th}
w^{\bullet}(g;x_1,\ldots , x_n)
:=\lim_{\Lambda\uparrow\mathbb R^{d}}w^{\bullet}_{\Lambda}(g;x_1,\ldots , x_n)
\end{equation}
and $w^{\bullet}_{\Lambda}$ is given in \eqref{act_with_points}.
Moreover, at infinite volume, we have the following bound:
\begin{equation}\label{boundforh}
\sup_{x_{1},\ldots,x_{n}\in\Lambda^{n}} \left| h^{(n)}(x_{1},\ldots,x_{n})
\right|\leq C.
\end{equation}
\end{theorem}

\subsection{Direct correlation function}\label{sec:direct}
Recalling \eqref{derstellandco} and the definition of $h^{(2)}_{\mathbb X}$ we observe that a new quantity (the first term in the right hand side) arises.
We will call it {\it direct correlation function}
and then we recognize in \eqref{derstellandco}
the Ornstein-Zernike equation \cite{OZ}:
\begin{equation}\label{OZv1}
h_{\mathbb X}^{(2)}(x_1, x_2)=c_{\mathbb X}(x_1,x_2)+ \int_{\mathbb X} c_{\mathbb X}(x_1,x_3) h_{\mathbb X}^{(2)}(x_3,x_2)\rho_{\mathbb X}^{(1)}(x_3)\, dx_3.
\end{equation}
This is particularly useful thanks to its ``renewal" property, namely that $h_{\mathbb X}^{(2)}(x_1, x_2)$
is divided into a ``direct" part denoted by $c_{\mathbb X}(x_1,x_3)$ up to some particle $x_3$ and an indirect one, in which we get again the ``renewed" $h_{\mathbb X}^{(2)}(x_3,x_2)$.
Equivalently, by functional differentiation it is easy to check (see \cite{stell1964}) that it can also be written as:
\begin{equation}\label{direct}
c_{\mathbb X}(x_1,x_2)=\frac{\delta}{\delta\rho(x_2)}\ln\frac{\delta \ln\Xi_{\mathbb X}}{\delta z(x_1)}.
\end{equation}
With this form, we can attempt to prove rigorously its power series expansion in terms of the density.

\begin{definition}\label{nodal}
We call a vertex {\it nodal} if there exists two white vertices in its connected component, which are different from the first vertex, such that all the paths between that pair of chosen white vertices passes through the first vertex. 
\end{definition}

For example, in the very definition of the
Ornstein-Zernike equation, if all involved quantities are represented by graphs, the vertex $x_3$ will be a nodal one.
We first prove the validity of the expansion of $\ln\frac{\delta \ln\Xi_{\mathbb X}}{\delta z(x_1)}$ following the previous results and obtaining $A$ as in \eqref{A}.
The next step is to do the
inversion in terms of $\rho$ 
which, according to Section~\ref{sec:inv}, it can be made by composing $A$ with $z=T(\rho)$
obtaining the 2-connected graphs as in \eqref{comp}.
The last step is to take $\frac{\delta}{\delta \rho(x_2)}$
which roots at a second (by now a ``$\rho$-type" vertex) $\rho(q_2)$.
The resulting graphs are 2-connected if we do not distinguish the colour of white/black.
It is instructive to compare this expansion 
to the procedure of getting $u^{(2)}_{\Lambda}$ in terms of $\rho$ since in the latter
we obtain the slightly different structure of articulation free graphs which allows for nodal vertices.
The difference is that for $h^{(2)}(x_1,x_2)$ we first take $\frac{\delta}{\delta z(x_2)}$ and then invert in terms of $\rho$, hence, by first pointing at $z(x_2)$, the ``final" nodal vertex
has survived and in the procedure of turning the $z$-vertices into $\rho$-vertices
it does not cancel as it does not ``hang off" (as it would have been the case in the first expansion where we switched the vertices into $\rho$'s before pointing at $x_2$). 

Alternatively, as in the previous subsection, we can avoid these intricate combinatorial
issues by working directly in the canonical ensemble. 
Recalling the definition of the set 
$\mathcal B_{2,n+2}$ we define the {\it direct correlation function}
in the canonical ensemble, i.e., for fixed volume $\Lambda$ and number of
particles $N+2$:
\begin{equation}\label{c2}
c^{(2)}_{\Lambda, N+2}(x_1,x_2):=\sum_{k=0}^N\frac{\rho^k}{k!}\sum_{g\in\mathcal B_{2,2+k}} w^{\bullet}_{\Lambda}(g; x_1,x_2).
\end{equation}
Then we have the following theorem \cite{kuna-tsagkaro2016}:

\begin{theorem}\label{thm3}
There exists a constant $c_0>0$ such that for all $\rho\,C(\beta)<c_0$,
the direct correlation function $c^{(2)}_{\Lambda, N+2}$ in \eqref{c2} 
converges in the thermodynamic limit,  to 
\begin{equation}\label{c2th}
c^{(2)}(x_1,x_2):=\sum_{k=0}^\infty\frac{\rho^k}{k!}\sum_{g\in\mathcal B_{2,2+k}} w^{\bullet}(g; x_1,x_2),
\end{equation} 
which is an analytic function in $\rho$, for $\rho\,C(\beta)<c_0$
and $w^{\bullet}$ is given in \eqref{act_with_points_th}.
The series \eqref{c2th} converges in the following sense:
\begin{equation}\label{boundforCbullet}
\sup_{x_1\in\Lambda}\int_\Lambda \frac{\rho^k}{k!}
\left|
\sum_{g\in\mathcal B_{2,2+k}} w^{\bullet}_{\Lambda}(g; x_1,x_2)
\right| \dd x_2\leq C e^{-c k},
\end{equation}
uniformly in $\Lambda$.

Furthermore, the direct correlation function $c^{(2)}_{\Lambda, N+2}$ in \eqref{c2} fulfils the Ornstein-Zernike equation \eqref{OZv1} up to the order $O(1/|\Lambda |$) and the limit function fulfils the Ornstein-Zernike equation exactly.
\end{theorem}
\begin{remark}
The constant $c_0$ in the above theorem is independent of the test function $\phi$, hence it is different from the constant $c_0$ in Theorem~\ref{thm1}. However it is determined in a similar way and can be estimated explicitly.
Moreover, as a direct consequence of \eqref{boundforCbullet}, we have that
\begin{equation}\label{boundsforOZ}
\sup_{x_{1}\in\Lambda}\int_{\Lambda}|c_{\Lambda,N}^{(2)}(x_{1},x_{2})| \dd x_{2}<\infty,
\end{equation}
which, together with \eqref{boundforh} (for $n=2$),
proves that the Ornstein-Zernike equation \eqref{OZv1} is well defined.
This is a key point in the proof and for the details we refer to \cite{kuna-tsagkaro2016}.
\end{remark}

\subsection{Inversion with respect to the interaction potential}

In Section~\ref{sec:one_point} we developed the mathematical machinery
that allows to prove the convergence of
expansions for thermodynamic quantities expressed via functional differentiation
(and not just scalar).
Based on that we would like
to investigate its applicability to the next order case: instead of differentiating with respect
to a function indexed by one label, to differentiate with respect to a function indexed by two labels and try to obtain a proof for the convergence of the inversion. The main example
is the
inversion of the two-point truncated correlation $h^{(2)}(x_1,x_2)$ with respect to
the pair potential $V(x_1,x_2)$ or a function of it, e.g. $f(x_1,x_2)$.
Establishing such an expansion would be very helpful in inverse problems, namely
computing the microscopic potential by knowing the correlations as given either by simulation or even $x$-ray experiments. This is also related to the inverse or realizability or moment problem, see \cite{KLS, ik20, ikr14} and the references therein.
Following \cite{morita-hiroike3}, Section~4 we have:
\begin{eqnarray}\label{twopointagain}
\rho_{\mathbb X}^{(2)}(x_1,x_2; z) & = & \frac{1}{\Xi_{\mathbb X}}z(x_1) z(x_2)\frac{\delta^2 \Xi_{\mathbb X}}{\delta z(x_1)\delta z(x_2)}\nonumber\\
& = & -\frac{2}{\beta}\frac{\delta\ln\Xi_{\mathbb X}}{\delta V(x_1,x_2)}\nonumber\\
& = & 2(1+f(x_1,x_2))\frac{\delta\ln\Xi_{\mathbb X}}{\delta f(x_1,x_2)}.
\end{eqnarray}
We observe that the above formula has a very similar structure with \eqref{eq:difflog}.
However, the fact that here instead of pointing at a vertex $x$, we have to point at the link $f(x_i,x_j)$ makes the previous machinery not directly applicable. Nevertheless, there are some
results in this direction following similar operations. For example, one would need to point at a link and see what happens when we remove it. We cite \cite{brydges_leroux} 
for the corresponding 
procedure and we hope to address this in a forthcoming work.

\subsection{Closures: Percus-Yevick equation}

We conclude this presentation by recalling that one of the aims of this review is
to connect to the existing literature and
ongoing research in liquid state theory where all these expansions have been 
extensively used, even in regimes in which we cannot prove their validity.
Being infinite power series, it is tempting to consider truncations and suggest approximate
numerical or analytical schemes.
For this to be successful one should have an idea of the absolute convergence 
of the series for distinguishing the dominant terms.
Furthermore, in some cases, these truncations lead to closed equations amenable to
computations.
Hence, the goal of this last section is to place the above results in the context of attempting
to identify, validate and compare different closures.
As a general conclusion, we understand that the tools used to prove convergence
are too crude to be able to distinguish between the various closures which are sometimes
designed on an empirical basis.
Despite this fact, it is an open territory for rigorous research using tools from analysis, probability and combinatorics for the investigation of denser regimes.

To start, we recall that
the Ornstein-Zernike equation \eqref{OZv1} is not a closed equation as it involves both
correlation functions $h_{\mathbb X}^{(2)}(q_1,q_2)$ and $c_{\mathbb X}(q_1,q_2)$.
One suggestion for a closure is the Percus-Yevick (PY) equation \cite{PY} that we describe 
below.
Starting from the OZ equation for $h_{\mathbb X}^{(2)}(r)$ and $c_{\mathbb X}(r)$,
following \cite{SPY}, one first introduces a new function $t$ as follows:
\begin{equation}\label{OZ}
 t_{\mathbb X}(r) : =c_{\mathbb X}*h_{\mathbb X}^{(2)}(r),
\end{equation}
where we use the convolution: $c_{\mathbb X}*h_{\mathbb X}^{(2)}(r) := \int c_{\mathbb X}(r')h_{\mathbb X}^{(2)}(r-r') \rho^{(1)}_{\mathbb X}(dr')$.
Then the OZ equation takes the form
\begin{equation}\label{OZ4}
h_{\mathbb X}^{(2)}(r)=c_{\mathbb X}(r)+t_{\mathbb X}(r).
\end{equation}
Note that all involved functions ($h_{\mathbb X}^{(2)}$, $c_{\mathbb X}$ and $t_{\mathbb X}$) are analytic functions in $\rho$.
Furthermore, $c_{\mathbb X}(r)$ can be written as
\begin{equation}\label{cofr}
c_{\mathbb X}(r)=f(r)(1+t_{\mathbb X}(r))+m_{\mathbb X}(r),
\end{equation}
where $f(r):=e^{-\beta V(r)}-1$ is a known function of the potential $V(r)$.
The relation \eqref{cofr} is essentially the definition of $m_{\mathbb X}(r)$ which is an analytic function of $\rho$ as well. Following \cite{SPY} the function $m_{\mathbb X}$  
can be expressed as a sum over two connected graphs which upon removal of the direct link $f$ connecting
the white vertices (if it is present) it is 2-connected (no articulation and no nodal points). 
For example, the first term of $m_{\mathbb X}(r)$ is the graph $1-3-2-4-1$.
However, in \cite{SPY}, {\it``the manipulations involved in obtaining these
infinite sums ... have been carried out in a purely formal way and we have not examined 
the important but difficult questions of convergence and the legitimacy of the
rearrangement of terms"}.
In the present review we showed how
to establish this convergence
with respect to $f$-bonds.
The convergence allows to quantify the error after truncating these terms. For example, $m$ is of the order $\rho^2$.
Furthermore, a future plan is to investigate whether
another suggestion could be made, relating some of the terms
in $m_{\mathbb X}(r)$ with respect to $t_{\mathbb X}(r)$, or by introducing another function (instead of $t(r)$)
as a candidate for a good choice for ``closing" OZ equation.
Combining \eqref{OZ} with \eqref{OZ4} and \eqref{cofr} we obtain:
\begin{equation}\label{combined}
t_{\mathbb X}=[f(1+t_{\mathbb X})+m_{\mathbb X}]*[f(1+t_{\mathbb X})+m_{\mathbb X}]
+[f(1+t_{\mathbb X})+m_{\mathbb X}]*t_{\mathbb X}.
\end{equation}
One version of PY equation is setting $m(r)\equiv 0$ and obtaining
a closed equation for $t_{\mathbb X}(r)$.

Alternatively, using \eqref{OZ4} and \eqref{cofr}  one can introduce the functions $y_{\mathbb X}(r)$ and $d_{\mathbb X}(r)$ by
\begin{equation}\label{yandd}
g_{\mathbb X}^{(2)}(r) = e^{-\beta V(r)} (1+t_{\mathbb X}(r)) +m_{\mathbb X}(r)=:e^{-\beta V(r)} y_{\mathbb X}(r) ,\qquad y_{\mathbb X}(r)=:1+t_{\mathbb X}(r)+d_{\mathbb X}(r),
\end{equation}
and hence $m_{\mathbb X}(r)=e^{-\beta V} d_{\mathbb X}(r)$. Thus,
we can rewrite \eqref{combined} as
\begin{eqnarray}\label{combined2}
y_{\mathbb X} & = & 1+ d_{\mathbb X} + [f\ y_{\mathbb X}+d_{\mathbb X}]*[e^{-\beta V}y_{\mathbb X} -1].
\end{eqnarray}
Again, setting $d_{\mathbb X}(r)\equiv 0$ we obtain another version of the PY equation. All
involved functions are analytic in $\rho$ and our results imply that the formal order in $\rho$ of $d_{\mathbb X}$ coincides with the actual order.
Now, one can investigate a method of systematically improving the PY equation, by adding 
some terms from $d_{\mathbb X}$ (or 
from $m$ for hard-core potentials).
For example, in \cite{SPY}  it was suggested to set $d_{\mathbb X}$ equal to the first order term in its expansion, 
since this gives
a {\it ``PY approximation that it leads to an approximate $g_{\mathbb X}$ that is exact
through terms of order $\rho^2$ in its virial expansion"}.
 A partial goal of the analysis in the present paper is to
provide a framework in which one can further investigate
such closure schemes and estimate the relevant error.
Other closures include the {\it Hypernetted Chain} (HNC) equation, the Born-Green-Yvon (BGY) hierarchy and the Kirkwood superposition
among many others. We think that it would be worthwhile to understand if there is some particular feature in anyone of these closures that
may allow to prove convergence in denser regimes.

\medskip\noindent
\subsubsection*{Acknowledgments} 
This is a review article to celebrate Errico Presutti's 80th birthday. It contains
research results over the last ten years starting with \cite{pulvirenti-tsagkaro2012} 
during my post-doc with him at Tor Vergata. As evident from the references, it is based on joint work and discussions with many colleagues and this research has strongly benefited from the fertile ground in the statistical mechanics group in Rome and Errico's guidance. I would also like to acknowledge many inspiring discussions with David Brydges, Marzio Cassandro, 
Roman Koteck\'y and Joel Lebowitz over all these years.

%

\end{document}